\documentclass[final]{acmtrans2e}

\usepackage{latexsym}

\newtheorem{theorem}{Theorem}[section]

\newtheorem{corollary}[theorem]{Corollary}
\newtheorem{proposition}[theorem]{Proposition}
\newtheorem{lemma}[theorem]{Lemma}
\newdef{definition}[theorem]{Definition}
\newdef{remark}[theorem]{Remark}

\newtheorem{exa}[theorem]{Example}

\newtheorem{exe}{Exercise}

\newcommand{\la}{\langle}
\newcommand{\ra}{\rangle}

\newcommand{\rrarrow}{\longrightarrow}

\long\def\comment#1{}

\title{A compositional Semantics for CHR}

\author{Maurizio Gabbrielli\\
  Universit\`{a} di Bologna
  \and
  Maria Chiara Meo\\
  Universit\`a ``G. D'Annunzio'' di Chieti-Pescara}

\begin{abstract}
Constraint Handling Rules (CHR) are a committed-choice declarative
language which has been designed for writing constraint solvers. A
CHR program consists of multi-headed guarded rules which allow one
to rewrite constraints into simpler ones until a solved form is
reached.

CHR has received a considerable attention, both from the practical
and from the theoretical side. Nevertheless, due the use of
multi-headed clauses, there are several aspects of the CHR
semantics which have not been clarified yet. In particular, no
compositional semantics for CHR has been defined so far.

In this paper we introduce a fix-point semantics which
characterizes the input/output behavior of a CHR program and which
is and-compositional, that is, which allows to retrieve the
semantics of a conjunctive query from the semantics of its
components. Such a semantics can be used as a basis to define
incremental and modular analysis and verification tools.
\end{abstract}

\begin{document}

            \begin{bottomstuff}
             Author's address:
             \newline
             Maurizio Gabbrielli,
  Dipartimento di Scienze dell'Informazione, Mura A. Zamboni 7, 40127 Bologna, Italy.
  {\tt gabbri@cs.unibo.it.}
  \newline Maria Chiara Meo, Dipartimento di Scienze, Viale Pindaro 42,
  65127 Pescara, Italy.
{\tt cmeo@unich.it.} \end{bottomstuff}

\maketitle

\section{Introduction}\label{sec:intro}

Constraint Handling Rules (CHR) \cite{Fr91,Fr94} are a
committed-choice declarative language which has been specifically
designed for writing constraint solvers. The first constraint
logic languages used mainly built-in constraint solvers designed
by following a ``black box'' approach. This made it hard to
modify, debug, and analyze a specific solver. Moreover, it was
very difficult to adapt an existing solver to the needs of some
specific application, and this was soon recognized as a serious
limitation since often practical applications involve application
specific constraints.

By using CHR one can easily introduce specific user-defined
constraints and the related solver into an host language. In fact,
a CHR program consists of (a set of) multi-headed guarded
simplification and propagation rules which are specifically
designed to implement the two most important operations involved
in the constraint solving process: Simplification rules allow to
replace constraints by simpler ones, while preserving their
meaning. Propagation rules are used to add new redundant
constraints which do not modify the meaning of the given
constraint and which can be useful for further reductions. It is
worth noting that the presence of multiple heads in CHR is an
essential feature which is needed in order to define reasonably
expressive constraint solvers (see the discussion in \cite{Fr94}).
However, such a feature, which differentiates this proposal from
many existing committed choice logic languages, complicates
considerably the semantics of CHR, in particular it makes very
difficult to obtain a compositional semantics, as we argue below.
This is unfortunate, as compositionality is an highly desirable
property for a semantics. In fact, a compositional semantics
provides the basis to define incremental and modular tools for
software analysis and verification, and these features are
essential in order to deal with partially defined components.
Moreover, in some cases, modularity allows to reduce the
complexity of verification of large systems by considering
separately smaller components.

In this paper we introduce a fix-point semantics for CHR which
characterizes the input/output behavior of a program and which is
and-compositional, that is, which allows to retrieve the semantics
of a conjunctive query from the semantics of its components.

In general, due to the presence of synchronization mechanisms,
the input/ouput semantics is not compositional for committed
choice logic languages and for most concurrent languages in
general. Indeed, the need for more complicate semantic structures
based on traces was recognized very early as a necessary condition
to obtain a compositional model, first for dataflow languages
\cite{Jo85} and then in the case of many other paradigms,
including imperative concurrent languages \cite{Br93} and
concurrent constraint and logic languages \cite{BP90c}.

When considering CHR this basic problem is further complicated:
due to the presence of multiple heads, the traces consisting of
sequences of input/ouput pairs, analogous to those used in the
above mentioned works, are not sufficient to obtain a
compositional semantics. Intuitively the problem can be stated as
follows. A CHR rule $r@\ h,g  \Leftrightarrow C \mid B$  cannot be
used to rewrite a goal $h$, no matter how the variables are
constrained (that is, for any input constraint), because the goal
consists of a single atom $h$ while the head of the rule contains
two atoms $h,g$. Therefore, if we considered a semantics based on
input/ouput traces, we would obtain the empty denotation for the
goal $h$ in the program consisting of the rule $r$ plus some rules
defining $B$. Analogously for the goal $g$. On the other hand, the
rule $r$ can be used to rewrite the goal $h,g$. Therefore,
provided that the semantics of $B$ is not empty, the semantics of
$h,g$ is not empty and cannot be derived from the semantics of $h$
and $g$, that is, the semantics is not compositional. It is worth
noting that even restricting to a more simple notion of
observable, such as the results of terminating computations, does
not simplify this problem. In fact, differently from the case of
ccp (concurrent constraint programming) languages, also the
semantics based on these observables (usually called resting
points) is not compositional for CHR. We have then to use some
additional information which allows us to describe the behavior of
goals in any possible and-composition without, of course,
considering explicitly all the possible and-compositions.

Our solution to obtain a compositional model is to use an
augmented semantics based on traces which includes at each step
two ``assumptions'' on the external environment and two
``outputs'' of the current process: Similarly to the case of the
models for ccp, the first assumption is made on the constraints
appearing in the guards of the rules, in order to ensure that
these are satisfied and the computation can proceed. The second
assumption is specific to our approach and contains atoms which
can appear in the heads of rules. This allows us to rewrite a goal
$G$ by using a rule whose head $H$ properly contains $G$: While
this is not possible with the standard CHR semantics, we allow
that by assuming that the external environment provides the
``difference'' $H$ minus $G$ and by memorizing such an assumption.
The first output element is the constraint produced by the
process, as usual. We also memorize at each step a second output
element, consisting of those atoms which are not rewritten in the
current derivation and which could be used to satisfy some
assumptions (of the second type) when composing sequences
representing different computations. Thus our model is based on
sequences of quadruples, rather than of simple input/output pairs.

Our compositional semantics is obtained by a fixpoint construction
which uses an enhanced transitions system implementing the rules
for assumptions described above. We prove the correctness of the
semantics w.r.t. a notion of observables which characterizes the
input/ouput behavior of terminating computations where the
original goal has been completely reduced to built-in constraints.
We will discuss later the extensions needed in order to
characterize different notions of results, such as the ``qualified
answers'' used in \cite{Fr94}.

The remaining of this paper is organized as follows. Next section
introduces some preliminaries about CHR and its operational
semantics. Section \ref{sec:compo} contains the definition of the
compositional semantics, while section \ref{sec:results} presents
the compositionality and correctness results. Section
\ref{sec:refined} discuss briefly a possible extension of this
work while section \ref{sec:concl} concludes by indicating
directions for future work.

\section{Preliminaries}

In this section we first introduce some preliminary notions  and
then define the CHR syntax and operational semantics. Even though
we try to provide a self-contained exposition, some familiarity
with constraint logic languages and first order logic could be
useful.

We first need to distinguish the constraints handled by an
existing solver, called built-in (or predefined) constraints, from
those defined by the CHR program, user-defined (or CHR)
constraints. An atomic constraint is a first-order predicate
(atomic formula). By assuming to use two disjoint sorts of
predicate symbols we then distinguish built-in atomic constraints
from CHR atomic constraints. A built-in constraint $c$ is defined
by

$$ c ::= a \ | \ c \wedge c \ | \ \exists_x a $$

where $a$ is an atomic built-in constraint \footnote{We could
consider more generally first order formulas as built-in
constraints, as far as the results presented here are concerned.}.
For built-in constraints we assume given a theory CT which
describes their meaning.

On the other hand, according to the usual CHR syntax,  we assume
that a user-defined constraint is a conjunction of atomic
user-defined constraints. We use $c,d$ to denote built-in
constraints, $g,h,k$ to denote CHR constraints and $a,b$ to denote
both built-in and user-defined constraints (we will call these
generically constraints). The capital versions of these notations
will be used to denote multisets of constraints. Furthermore we
denote by ${\cal U}$ the set of user-defined constraints and by
${\cal B}$ the set of built-in constraints.

We will often use ``,'' rather than $\wedge$ to denote conjunction
and we will often consider a conjunction of atomic constraints as
a multiset of atomic constraints. In particular, we will use this
notation based on multisets in the syntax of CHR. The notation
$\exists_{-V}\phi$, where $V$ is a set of variables, denotes the
existential closure of a formula $\phi$ with the exception of the
variables $V$ which remain unquantified. $Fv(\phi)$ denotes the
free variables appearing in $\phi$ and we denote by $\cdot$ the
concatenation of sequences and by $\varepsilon$ the empty
sequence. Furthermore $\uplus$ denotes the multi-set union, while
we consider $\setminus$ as an overloaded operator used both for
set and multi-set difference (the meaning depends on the type of
the arguments).

We are now ready to introduce the CHR syntax as defined in \cite{Fr94}.

\begin{definition}\label{Syntax}[Syntax]
A CHR {\em simplification} rule has the form \[{\it r}@ H
\Leftrightarrow C \mid B\] while a CHR propagation rule has the
form \[{\it r}@ H \Rightarrow C \mid B, \]  where ${\it r }$ is a
unique identifier of a rule, $H$ is a multiset of user-defined
constraints, $C$ is a multiset of built-in constraints and $B$ is
a possibly empty multi-set of (built-in and user-defined)
constraints\footnote{Some papers consider also simpagation rules,
Since these are abbreviations for propagation and simplification
rules we do not need to introduce them.}. A CHR program is a
finite set of CHR simplification and propagation rules.
\end{definition}

We prefer to use multisets rather than sequences
(as in the original CHR papers) since multisets appear to
correspond more precisely to the nature of CHR rules.
Moreover in this paper we will not use the
identifiers of the rules, which will then
be omitted.

A CHR goal is a multiset of (both user-defined an built-in)
constraints. $Goals$ is the set of all goals.

We describe now the operational semantics of CHR as provided by
\cite{Fr94} by using a transition system $T_{s}= ({\it Conf_{s}},
\rrarrow_{s})$ ($_{s}$ here stands for ``standard'', as opposed to
the semantics we will use later). Configurations in ${\it
Conf_{s}}$ are triples of the form $\langle G,K,d \rangle$ where
$G$ are the constraints that remain to be solved, $K$ are the
user-defined constraints that have been accumulated and $d$ are
the built-in constraints that have been simplified\footnote{In
\cite{Fr94} triples of the form $\langle G,K,d \rangle_{\cal V}$
were used, where the annotation ${\cal V}$, which is not changed
by the transition rules, is used to distinguish the variables
appearing in the initial goal from the variables which are
introduced by the rules. We can avoid such an indexing by
explicitly referring to the original goal.}.

\begin{table*}[tbp]
\begin{center}
\begin{tabular}{|lll|}  \hline
&\mbox{   }&\mbox{   }
\\
\mbox{\bf Solve}& $\frac {\displaystyle   CT \models c \wedge d
\leftrightarrow d' \hbox{ and c is a built-in constraint}}
{\displaystyle \la(c,G), K,d\ra \rrarrow_{s} \la G, K,d'\ra
}$&\mbox{ }
\\
&\mbox{   }&\mbox{   }
\\
\mbox{\bf Introduce}& $\frac {\displaystyle  \hbox{h is a
user-defined constraint}} {\displaystyle \la(h ,G), K,d\ra
\rrarrow_{s} \la G, (h,K),d\ra }$&\mbox{ }
\\
&\mbox{   }&\mbox{   }
\\
&\mbox{   }&\mbox{   }
\\
\mbox{\bf Simplify}& $\frac {\displaystyle   H \Leftrightarrow C
\mid B \in P \ \ \  x = Fv(H) \ \ \ \ CT \models d \rightarrow
\exists _{x} ((H=H')\wedge C)}  {\displaystyle \la G,H'\wedge
K,d\ra \rrarrow_{s} \la B\wedge G,K, H=H'\wedge d \ra }$&\mbox{ }
\\
&\mbox{   }&\mbox{   }
\\
\mbox{\bf Propagate}& $\frac {\displaystyle   H \Rightarrow C \mid
B \in P \ \ \  x = Fv(H) \ \ \ \ CT \models d \rightarrow \exists
_{x} ((H=H')\wedge C) } {\displaystyle \la G,H'\wedge K,d\ra
\rrarrow_{s} \la B\wedge G,H'\wedge K, H=H'\wedge d \ra }$&\mbox{
}
\\
&\mbox{   }&\mbox{   }
\\
\hline
\end{tabular}
\caption{The standard transition system for CHR}\label{tor}
\end{center}
\end{table*}

An {\em initial configuration} has the form
\[\langle G,\emptyset,\emptyset \rangle\]
and consists of a goal $G$, an empty user-defined constraint and an empty built-in constraint.

A {\em final configuration} has either the form
\[\langle G,K, \tt false \rangle,\]
when it is {\em failed}, i.e. when it contains  an inconsistent
built-in constraint store represented by the unsatisfiable
constraint ${\tt false}$, or has the form
\[\langle \emptyset,K,d \rangle\] when it is successfully terminated since there are
no applicable rules.

Given a program $P$, the transition relation $\rrarrow_{s}
\subseteq {\it Conf} \times {\it Conf}$ is the least relation
satisfying the rules  in Table \ref{tor} (for the sake of
simplicity, we omit indexing the relation with the name of the
program). The {\bf Solve} transition allows to update the
constraint store by taking into account a built-in constraint
contained in the goal. Without loss of generality, we will assume
that $Fv(d') \subseteq Fv(c) \cup Fv(d)$. The {\bf Introduce}
transition is used to move a user-defined constraint from the goal
to the CHR constraint store, where it can be handled by applying
CHR rules.  The transitions {\bf Simplify} and {\bf Propagate}
allow to rewrite user-defined constraints (which are in the CHR
constraint store) by using rules from the program. As usual, in
order to avoid variable names clashes, both these transitions
assume that clauses from the program are renamed apart, that is
assume that all variables appearing in a program clause are fresh
ones. Both the {\bf Simplify} and {\bf Propagate} transitions are
applicable when the current store ($d$) is strong enough to entail
the guard of the rule ($c$), once the parameter passing has been
performed (this is expressed by the equation $H=H'$). Note that,
due to the existential  quantification over the variables $x$
appearing in $H$, in such a parameter passing the information flow
is from the actual parameters (in $H'$) to the formal parameters
(in $H$), that is, it is required that the constraints $H'$ which
have to be rewritten are an instance of the head $H$. When
applied, both these transitions add the body $B$ of the rule to
the current goal and the equation $H=H'$, expressing the parameter
passing mechanism, to the built-in constraint store. The
difference between {\bf Simplify} and {\bf Propagate} is in the
fact that while the former transition removes the constraints $H'$
which have been rewritten from the CHR constraint store, this is
not the case for the latter.

Given a goal $G$, the operational semantics that we consider
observes the final stores of computations terminating with an
empty goal and an empty user-defined constraint. We call these
observables data sufficient answers following the therminology of
\cite{Fr94}.

\begin{definition}\label{opsemsa}[Data sufficient answers]
Let $P$ be a program and let $G$ be a goal.  The set ${\cal
SA}_{P}(G)$ of data sufficient answers for the query $G$ in the
program $P$ is defined as follows
\[{\cal SA}_P(G) = \begin{array}[t]{l}  \{
\la \exists_{-Fv(G)}  d \ra \mid \la G, \emptyset,\emptyset \ra
\rrarrow_{s} ^*\la \emptyset, \emptyset, d \ra \not \rrarrow_s \}.
\\
\cup
\\
\{ \la {\tt false} \ra \mid \la G, \emptyset,\emptyset \ra
\rrarrow_{s} ^*\la G', K, {\tt false} \ra   \}.
\end{array}
\]
\end{definition}

In \cite{Fr94} it is also considered the following different
notion of answer, obtained by computations terminating with a
user-defined constraint which does not need to be empty.

\begin{definition} \label{opsemqa}[Qualified answers]
Let $P$ be a program and let $ G$ be a goal. The set ${\cal
QA}_{P}(G)$ of qualified answers for the query $G$ in the program
$P$ is defined as follows
\[{\cal QA}_P(G) =\begin{array}[t]{l} \{ \la \exists_{-Fv(G)} K\wedge d \ra \mid \la G,
\emptyset,\emptyset \ra \rrarrow_{s} ^*\la \emptyset, K, d \ra
\not \rrarrow_s \}
\\
\cup
\\
\{ \la {\tt false} \ra \mid \la G, \emptyset,\emptyset \ra
\rrarrow_{s} ^*\la G', K, {\tt false} \ra   \}.
\end{array}
\]
\end{definition}

We discuss in Section \ref{sec:concl} the extensions needed to
characterize also qualified answers. Note that both previous
notions of observables characterize an input/output behavior,
since the input constraint is implicitly considered in the goal.

In the remaining of this paper we will consider only
simplification rules since propagation rules can be mimicked by
simplification rules, as far as the results contained in this
paper are concerned.

Note that in presence of propagation rules the ``naive''
operational semantics that we consider in this paper introduces
redundant infinite computations: Since propagation rules do not
remove user defined constraints (see rule Propagate in Table
\ref{tor}), when a propagate rule is applied it introduces an
infinite computation (obtained by subsequent applications of the
same rule). Note however that this does not imply that in presence
of an active propagation rule the semantics that we consider are
empty. In fact, the application of a simplification rule after a
propagation rule can cause the termination of the computation, by
removing the atoms which are needed by the head of the propagation
rule. It is also possible to define a more refined operational
semantics (see \cite{Ab97} and \cite{DGS04}) which avoids these
infinite computations by allowing to apply at most once a
propagation rule to the same constraints. We discuss in Section
\ref{sec:refined} the modifications needed in our construction to
take into account this more refined semantics.

\section{A compositional trace semantics}\label{sec:compo}

Given a program $P$, we say that a semantics ${\cal S}_P$ is
and-com\-po\-si\-tio\-nal if ${\cal S}_P(A,B) = {\cal C}({\cal
S}_P(A),{\cal S}_P(B))$ for a suitable composition operator ${\cal
C}$ which does not depend on the program $P$. As mentioned in the
introduction, due to the presence of multiple heads in CHR, the
semantics which associates
to a program $P$ the function ${\cal SA}_{P}$ is not
and-compositional, since goals which have the same input/ouput
behavior can behave differently when composed with other goals.
Consider for example the program $P$ consisting of the single rule
$$
g,h \Leftrightarrow true | c
$$
(where $c$ is a built-in constraint). According to Definition
\ref{opsemqa} we have that  ${\cal SA}_P(g) = {\cal SA}_{P}(k) =
\emptyset$, while $${\cal SA}_P(g,h)= \{ \la \exists_{-Fv(g,h)} c
\ra \} \neq \emptyset = {\cal SA}_{P}(k,h).$$ An analogous example
can be made to show that also the semantics ${\cal QA}$ is not
and-compositional.

The problem exemplified above is different from the classic
problem of concurrent languages where the interaction of
non-de\-ter\-mi\-nism and synchronization makes the input/output
observables non-com\-po\-si\-tio\-nal. For this reason,
considering simply sequences of (input-output) built-in
constraints is not sufficient to obtain a compositional semantics
for CHR. We have to use some additional information which allows
us to describe the behavior of goals in any possible
and-composition without, of course, considering explicitly all the
possible and-compositions.

The basic idea of our approach is to collect in the semantics also
the ``missing'' parts of heads which are needed in order to
proceed with the computation. For example, when considering the
program $P$ above, we should be able to state that the goal $g$
produces the constraint $c$, provided that the external
environment (i.e. a conjunctive goal) contains the user-defined
constraint $h$. In other words, $h$ is an assumption which is made
in the semantics describing the computation of $g$. When composing
(by using a suitable notion of composition) such a semantics with
that one of a goal which contains $h$  we can verify that the
``assumption'' $h$ is satisfied and therefore obtain the correct
semantics for $g,h$. In order to model correctly the interaction
of different processes we have to use sequences, analogously to
what happens with other concurrent paradigms.

This idea is developed by defining a new transition system which
implements this mechanism based on assumptions for dealing with
the missing parts of heads. The new transition system allows one
to generate the sequences appearing in the compositional model by
using a standard fix-point construction. As a first step in our
construction we modify the notion of configuration used before:
Since we do not need to distinguish user-defined constraints which
appear in the goal from the user-defined constraints which have
been already considered for reduction, we merge the first and the
second components of previous triples. Thus we do not need anymore
{\bf Introduce} rule. On the other hand, we need the information
on the new assumptions, which is added as a label of the
transitions.

\begin{table*}[tbp]
\begin{center}
\begin{tabular}{|lll|}  \hline
&\mbox{   }&\mbox{   }
\\
\mbox{\bf Solve'}& $\frac {\displaystyle CT \models c  \wedge d
\leftrightarrow d'}  {\displaystyle \la c \wedge G, d\ra
\rrarrow_P^\emptyset \la G, d'\ra }$&\mbox{ }
\\
&\mbox{   }&\mbox{   }
\\
\mbox{\bf Simplify'}& $\frac {\displaystyle  H \Leftrightarrow C
\mid B \in P \ \ \ \ x = Fv(H) \ \ \ \ G\neq \emptyset \ \ \ \ CT
\models d \rightarrow \exists _x ((H=(G,K)) \wedge C)}
{\displaystyle \la G\wedge A, d\ra \ \rrarrow_P^K \la
B^{i+1}\wedge A, d \wedge (H =(G,K)) \ra }$&\mbox{ }
\\
&\mbox{   }&\mbox{   }
\\
&\mbox{where } $i \mbox{ is the maximal index occurring in the
goal } G\wedge A$ &\mbox{ }
\\
&\mbox{   }&\mbox{   }
\\
\hline
\end{tabular}
\caption{The transition system for the compositional
semantics}\label{tcs}
\end{center}
\end{table*}

Thus we define a transition system $T= ({\it Conf}, \rrarrow_P)$
where configurations in {\it Conf } are pairs: the first component
is a multiset of indexed atoms (the goal) and the second one is a
built-in constraint (the store). Indexes are associated to atoms
in order to denote the point in the derivation where they have
been introduced. Atoms in the original goals are indexed by $0$,
while atoms introduced at the i-th derivation step are indexed by
$i$. Given a program $P$, the transition relation $\rrarrow_P
\subseteq{\it Conf} \times {\it Conf} \times \wp({\cal U}) $ is
the least relation satisfying the rules in Table~\ref{tcs} (where
$\wp(A)$ denotes the set consisting of all the subsets of $A$).
Note that we consider only {\bf Solve} and {\bf Simplify} rules,
as the other rules as previously mentioned are redundant in this
context. {\bf Solve'} is the same rule as before, while the {\bf
Simplify'} rule is modified to consider assumptions: When reducing
a goal $G$ by using a rule having head $H$, the multiset of
assumptions $K = H\setminus G$ (with $H\neq K$) is used to label
the transition ($\setminus$ here denotes multiset difference).
Indexes allow us to distinguish different occurrences of the same
atom which have been introduced in different derivation steps. We
will use the notation $G^{i}$ to indicate that all the atoms in
$G$ are indexed by $i$.

When indexes are not needed we will simply omit them. As before,
we assume that program rules to be used in the new simplify rule
use fresh variables to avoid names clashes.

The semantics domain of our compositional semantics is based on
sequences which represent derivations obtained by the transition
system in Table \ref{tcs}. More precisely, we first consider
``concrete'' sequences consisting of tuples of the form $\la
G,c,K,G',d\ra$: Such a tuple represents a derivation step $\la
G,c\ra \rrarrow_P^K \la G',d\ra$. The sequences we consider are
terminated by tuples of the form $\la G,c,\emptyset,G,c\ra$, which
represent a terminating step (see the precise definition below).
Since a sequence represents a derivation, we assume that the
``output'' goal $G'$ at step $i$ is equal to the ``input'' goal
$G$ at step $i+1$, that is, we assume that if
$$\ldots \la
G_i,c_i,K_i,G_i',d_i\ra \la G_{i+1},c_{i+1},K_{i+1},G'_{i+1},d_{i+1} \ra \ldots $$
appears in a sequence, then
$G_i' = G_{i+1}$ holds.

On the other hand, the input store $c_{i+1}$ can be different from
the output store $d_i$ produced at previous step, since we need to
perform all the possible assumptions on the constraint $c_{i+1}$
produced by the external environment in order to obtain a
compositional semantics. However, we assume that if
$$\ldots \la G_i,c_i,K_i,G_i',d_i\ra
\la G_{i+1},c_{i+1},K_{i+1},G'_{i+1},d_{i+1} \ra \ldots $$ appears
in a sequence then $CT \models c_{i+1} \rightarrow d_i$ holds:
This means that the assumption made on the external environment
cannot be weaker than the constraint store produced at the
previous step. This reflects the monotonic nature of computations,
where information can be added to the constraint store and cannot
be deleted from it. Finally note that assumptions on user-defined
constraints (label $K$) are made only for the atoms which are
needed to ``complete'' the current goal in order to apply a
clause. In other words, no assumption can be made in order to
apply clauses whose heads do not share any predicate with the
current goal.

The set of the above described ``concrete'' sequences,  which
represent derivation steps performed by using the new transition
system, is denoted by ${\cal S}eq$.

\noindent From these concrete sequences we extract some more
abstract sequences which are the objects of our semantic domain:
From each tuple $\la G,c,K,G',d\ra$ in a sequence $\delta\in {\cal
S}eq$ we extract  a tuple of the form $\la c,K,H,d\ra$ where we
consider as before the input and output store ($c$ and $d$,
respectively) and the assumptions ($K$), while we do not consider
anymore the output goal $G'$.  Furthermore, we restrict the input
goal $G$ to that part $H$ consisting of all those user-defined
constraints which will not be rewritten in the (derivation
represented by the) sequence $\delta$. Intuitively $H$ contains
those atoms which are available for satisfying assumptions of
other goals, when composing two different sequences (representing
two derivations of different goals). We also assume that if \[\la
c_i,K_i,H_i,d_i\ra \la c_{i+1},K_{i+1},H_{i+1},d_{i+1}\ra\] is in
a sequence then $H_i\subseteq H_{i+1}$ holds, since these atoms
which will not be rewritten in the derivation can only augment.
Finally, indexes are not used in the abstract sequences (they are
only needed to define stable atoms, see Definition
\ref{def:alpha}).

We then define formally the semantic domain as follows.

\begin{definition} \label{interpretation}[Abstract sequences]
The semantic domain ${\cal D}$ containing all the possible
(abstract) sequences is defined as the set
\[ \begin{array} {ll} {\cal D} = & \{\la c_1,K_1,H_1,d_1\ra \ldots \la
c_n,\emptyset,H_n,c_n\ra\mid\\
& \hspace*{0.2cm} \begin{array}[t]{ll}
\hbox{for each }j,\  1\leq j\leq n \hbox{ and for each } i, 1\leq i\leq n -1, \ \\
H_j \hbox{ and } K_i \hbox{ are multisets of CHR  (non indexed) constraints,}
\\
c_j, d_i  \hbox{ are built-in constraints and }
CT \models d_i \rightarrow c_i, \\
H_i\subseteq H_{i+1} \hbox{ and } CT \models c_{i+1} \rightarrow d_i \mbox{ holds } \}.
 \end{array}
 \end{array}
\]
\end{definition}

In order to define our semantics we need three more notions.
First, we define an abstraction operator $\alpha$ which extracts
from the concrete sequences in ${\cal S}eq$ (representing exactly
derivation steps) the abstract sequences used in our semantic
domain.

\begin{definition} \label{def:alpha}[Abstraction and Stable atoms]
Let \[\delta = \la G_1,c_1, K_1, G_2,d_1\ra \ldots \la G_n,
c_n,\emptyset,G_n,c_n\ra\] be a sequence of derivation steps where
we assume that atoms are indexed as previously specified. We say
that an indexed atom $A^j$ is stable in $\delta$ if $A^j$ appears
in $G_i$, for each $1\leq i\leq n$. The abstraction operator
$\alpha: {\cal S}eq \rightarrow {\cal D}$ is then defined
inductively as
\[\begin{array}{lll}
\alpha(\varepsilon) = \varepsilon
\\
\alpha(\la G,c,K,G',d\ra\cdot \delta') = \beta( \la c,K,H,d\ra)\cdot \alpha(\delta')
\end{array}
\]
where $H$ is the multiset consisting of all the atoms in $G$ which
are stable in $\la G,c,K,G',d\ra\cdot\delta'$ and the function
$\beta$ simply removes the indexes from the atoms in $H$.

\end{definition}

Then we need the notion of ``compatibility'' of a tuple w.r.t. a
sequence. To this aim we first provide some further notation:
Given a sequence $\delta$ of derivation steps
\[\la G_1,c_1, K_1, G_2, d_1 \ra \la G_2,
c_2, K_2, G_3, d_2 \ra \ldots \la G_{n}, c_n,
\emptyset,G_{n},c_n\ra\] we denote by $length(\delta)$ the length
of the derivation $\delta$ (i.e. the number of tuples in the
sequence). Moreover using $t$ as a shorthand for the tuple $\la
G_1,c_1, K_1, G_2, d_1 \ra$ we define

\begin{description}
    \item[] $V_{loc}(t)= Fv(G_2, d_1) \setminus Fv(G_1,c_1,
    K_1)$,
    \item[] $V_{ass}(\delta)=\bigcup_{i=1}^{n-1} Fv(K_i)$ (the
    variables in the assumptions of $\delta$),
    \item[] $V_{stable}(\delta)= Fv(G_n)$ (the variables
    in all the stable multisets of $\delta$),
    \item[] $V_{constr}(\delta)= \bigcup_{i=1}^{n-1} Fv(d_i) \setminus Fv(c_i)$
    (the variables in the output constraints of $\delta$ which are not
    in the corresponding input constraints) and
    \item[] $V_{loc}(\delta)= \bigcup_{i=1}^{n-1} Fv(G_{i+1}, d_i) \setminus Fv(G_i,c_i,
    K_i)$ (the local variables of $\delta$,
    namely the variables in the clauses used in the derivation $\delta$).
\end{description}

We then define the notion of compatibility as follows.

\begin{definition}\label{def:compatibility}
Let $t=\la G_1,c_1, K_1, G_2, d_1 \ra$  a tuple representing a
derivation step for the goal $G_1$ and let $\delta= \la G_2, c_2,
K_2, G_3, d_2 \ra \ldots \la G_{n}, c_n, \emptyset,G_{n},c_n\ra$
be a sequence of derivation steps for $G_2$. We say that $t$ is
compatible with $\delta$ if the following hold:
\begin{enumerate}
    \item $CT \models c_2 \rightarrow d_1$,
    \item $V_{loc}(\delta) \cap Fv (t) = \emptyset$,
    \item $V_{loc}(t)\cap V_{ass}(\delta) = \emptyset$ and
    \item for $i \in [2,n]$, $V_{loc}(t) \cap Fv(c_i) \subseteq
    \bigcup_{j=1}^{i-1}  Fv(d_j) \cup V_{stable}(\delta)$.
\end{enumerate}
\end{definition}
The first three condition reflect the monotonic nature of
computations, that the clauses in a derivation are renamed apart
and that the variables in the assumptions are disjoint from the
variables in the clauses used in a derivation. The last condition
ensure that the local variables in a derivation $\delta$ and in
the abstraction of $\delta$ are the same (see
Lemma~\ref{lem:stessevar}). Note that if $t$ is compatible with
$\delta$ then, by using the notation above, $ t \cdot \delta$ is a
sequence of derivation steps for $G_1$. We can now define the
compositional semantics.

\begin{definition}\label{def:compsem}[Compositional semantics]
Let $P$ be a program and let $G$ be a goal. The compositional
semantics of $G$ in the program $P$, ${\cal S}_P: Goals
\rightarrow \wp({\cal D})$, is defined as
\[
{\cal S}_P(G) = \alpha ( {\cal S}'_P(G))
\]
where $\alpha$ is the pointwise extension to sets of the operator
given in Definition \ref{def:alpha} and ${\cal S}'_P: Goals
\rightarrow \wp({\cal S}eq)$ is defined as follows:
\[
\begin{array}[t]{lll}
{\cal S'}_P(G) = & \{\la G,c,K,G',d\ra\cdot \delta \in {\cal S}eq
\mid & \begin{array}[t]{l}CT\not \models c\leftrightarrow {\tt
false}, \
 \la G, c \ra \rrarrow_P^K \la G', d \ra \\
 \mbox{
and } \delta \in {\cal S'}_P(G')  \hbox{ for some $\delta$ such
that
 } \\ \la G,c, K, G',d\ra \hbox{ is compatible with $\delta$} \}
\end{array}
\\
&\cup
\\
&\{ \la G,c,\emptyset,G,c\ra\in {\cal S}eq \}.
\end{array}
\]

Formally ${\cal S}'_P(G)$ is defined as the least fixed-point of
the corresponding operator $\Phi\in ({\it Goals}\rightarrow
\wp({\cal S}eq)) \rightarrow {\it Goals}\rightarrow \wp({\cal
S}eq)$ defined by
\[
\begin{array}[t]{lll}
\Phi(I)(G) = &
 \{ \la G,c, K, G',d\ra \cdot \delta  \in {\cal S}eq
\mid & \begin{array}[t]{l} CT\not \models c\leftrightarrow {\tt
false}, \
\la G, c \ra \rrarrow_P^K \la G', d \ra \\
 \mbox{ and } \delta \in I(G')
 \hbox{ for some $\delta$ such that } \\ \la G,c, K, G',d\ra \hbox{
is compatible with $\delta$} \}
\end{array}
\\
&\cup
\\
&\{\la G,c, \emptyset, G ,c\ra \in {\cal S}eq\}.
\end{array}
\]
\end{definition}

In the above definition, $I: {\it Goals}\rightarrow \wp({\cal
S}eq)$  stands for a generic interpretation assigning to a goal a
set of sequences, and the ordering on the set of interpretations
${\it Goals}\rightarrow \wp({\cal S}eq)$ is that of (point-wise
extended) set-inclusion.  It is straightforward to check that
$\Phi$ is continuous (on a CPO), thus standard results ensure that
the fixpoint can be calculated by $\sqcup_{n\geq 0} \phi^n
(\bot)$, where $\phi^0$ is the identity map and for $n>0$,
$\phi^n$ = $\phi \circ \phi^{n-1}$ (see for example \cite{DP90}).

\section{Compositionality and correctness}\label{sec:results}

In this section we prove that the semantics defined above is
and-compositional and correct w.r.t. the observables ${\cal
SA}_{P}$.

In order to prove the compositionality result we first need to
define how two sequences describing a computation of $A$ and $B$,
respectively, can be composed in order to obtain a computation of
$A,B$. Such a composition is defined by the (semantic) operator
$\parallel$ which performs an interleaving of the actions
described by the two sequences and then eliminates the assumptions
which are satisfied in the resulting sequence. For technical
reasons, rather than modifying the existing sequences, the
elimination of satisfied assumptions is performed on new sequences
which are generated by a closure operator $\eta$ defined as
follows.

\begin{definition}\label{def:normalization}
Let $W$ be a multiset of indexed atoms, $\sigma$ be a sequence in
${\cal D}$ of the form
$$\la c_1,K_1, H_1, d_1\ra\,
\la c_2,K_2, H_2, d_2\ra \ldots \la c_n, K_n, H_n, d_n\ra$$ and
let
\[\tilde H_1= H_1^1 \mbox{ and for $i\in [2,n]$ }
\tilde H_i=\tilde H_{i-1}  \uplus (H_i \setminus H_{i-1})^i,\]
where we use the notation $H^i$ to indicate that all the atoms in
$H$ are indexed by $i$ and $\setminus$ denotes the multisets
difference.

We denote by $\sigma\setminus W$ the sequence
$$\beta(\la c_1,K_1, \tilde H_1\setminus W, d_1\ra\,
\la c_2,K_2, \tilde H_2\setminus W, d_2\ra\ldots \la c_n, K_n,
\tilde H_n\setminus W, d_n\ra)$$ where the multisets difference
$\tilde H_i\setminus W$ considers indexes and, as in
Definition~\ref{def:alpha}, the function $\beta$ simply removes
the indexes from the stable atoms.

The operator $\eta: \wp({\cal D}) \rightarrow \wp({\cal D})$ is
defined as follows. Given $S\in \wp({\cal D})$, $\eta(S)$  is the
least set satisfying the following conditions:
\begin{enumerate}
    \item $S \subseteq \eta (S)$;
    \item if $\sigma'\cdot \la c,K, H, d\ra \cdot \sigma'' \in \eta (S)$
    then  $(\sigma' \cdot \la c, K\setminus K',H,d\ra\cdot\sigma'')
    \setminus W \in \eta (S)$
\end{enumerate}
where $K'=\{A_1, \ldots, A_n\}\subseteq K$ is a multiset such that
there exists a multiset of indexed atoms $W =\{B_1^{j_1}, \ldots,
B_n^{j_n}\}\subseteq \tilde H$ such that $CT\models c\wedge
B_l\leftrightarrow c \wedge A_l$, for each $l\in [1,n]$.
\end{definition}

A few explanations are in order. The operator $\eta$ is an upper
closure operator\footnote{$S \subseteq \eta (S)$ holds by
definition, and it is easy to see that $\eta(\eta(S))=\eta(S)$
holds and that $S \subseteq S'$ implies $\eta (S)\subseteq \eta
(S')$.} which saturates a set of sequences $S$ by adding new
sequences where redundant assumptions can be removed: an
assumptions $a$ (in $K_i$) can be removed if $a^j$ appears as a
stable atom (in $\tilde H_i$). Once a stable atom is ``consumed''
for satisfying an assumption it is removed from (the multiset of
stable atoms of) all the tuples appearing in the sequence, to
avoid multiple uses of the same atom. Note that stable atoms are
considered without the index in the condition $CT\models c\wedge
B_l\leftrightarrow c \wedge A_l$, while they are considered as
indexed atoms in the removal operation $\tilde H_i\setminus W$.
The reason for this slight complication is explained by the
following example. Assume that we have the set $S$ consisting of
the only sequence $\la c,\emptyset, \{a\},d\ra \la c',\{a\},
\{a,a\},d'\ra \la c'',\emptyset, \{a,a\},c''\ra$. From this
sequence, we construct a new one, where the stable atoms are
indexed as follows: \[\la c,\emptyset, \{a^1\},d\ra \la c',\{a\},
\{a^1,a^2\},d'\ra \la c'',\emptyset, \{a^1,a^2\},c''\ra.\] Such a
new sequence indicates that at the second step we have an
assumption $a$, while both at the first and at the second step we
have produced a stable atom $a$, which has been indexed by $1$ and
$2$, respectively. In order to satisfy the assumption $a$ we can
use either $a^1$ or $a^2$.\\
However, depending on what indexed
atom we use, we obtain two different simplified sequences in
$\eta(S)$, namely
\\
$\begin{array}{l}
  \la c,\emptyset, \emptyset ,d\ra \la c',\emptyset, \{a\},d'\ra\la
c'',\emptyset, \{a\},c''\ra \mbox{ and  }
  \la c,\emptyset, \{a\},d\ra \la
c',\emptyset, \{a\},d'\ra\la c'',\emptyset, \{a\},c''\ra,  \\
\end{array}
$\\ which describe correctly the two different situations. It is
also worth noting that it is possible to disregard indexes in the
result of the normalization operator

Before defining the composition operator $\parallel$ on sequences
we  need a notation for the sequences in ${\cal
D}$ analogous to that one introduced for sequences of derivation steps: \\
Let $\sigma= \la c_1, K_1, H_1,d_1\ra\la c_2, K_2, H_2,d_2\ra\cdots
\la c_n, \emptyset, H_n,d_n\ra \in {\cal D}$
be a sequence for the goal $G$. We define
\begin{description}
    \item[] $V_{ass}(\sigma)=\bigcup_{i=1}^{n-1} Fv(K_i)$ (the
    variables in the assumptions of $\sigma$),
    \item[] $V_{stable}(\sigma)= Fv(H_n)=\bigcup _{i=1}^n Fv(H_i)$ (the variables
    in the stable multisets of $\sigma$),
    \item[] $V_{constr}(\sigma)= \bigcup_{i=1}^{n-1} Fv(d_i) \setminus Fv(c_i)$
    (the variables in the output constraints of $\sigma$ which are not
    in the corresponding input constraints),
    \item[] $V_{loc}(\sigma)= (V_{constr}(\sigma)\cup
    V_{stable}(\sigma))\setminus(V_{ass}(\sigma)\cup Fv(G))$
    (by using Condition 4 of Definition \ref{def:compatibility} and by
    Lemma~\ref{lem:stessevar},
    the local variables of a sequence $\sigma$ are the local
    variables of the derivations $\delta$ such $\alpha(\delta)=\sigma$).
\end{description}

We can now define the composition operator $\parallel$  on
sequences. To simplify the notation we denote by $\parallel$ both
the operator acting on sequences and that one acting on sets of
sequences.

\begin{definition}\label{def:composition}
The operator $\parallel: {\cal D}\times {\cal D} \rightarrow
\wp({\cal D})$ is defined inductively as follows. Assume that
$\sigma_1 = \la c_1, K_1, H_1,d_1\ra\cdot \sigma'_1$ and $\sigma_2
= \la c_2, K_2, H_2,d_2\ra\cdot \sigma'_2$ are sequences for the
goals $G_1$ and $G_2$, respectively. If
\begin{eqnarray}\label{eqnuno}
(V_{loc}(\sigma_1)\cup
Fv(G_1) )\cap ( V_{loc}(\sigma_2) \cup Fv(G_2))= Fv(G_1) \cap
Fv(G_2)
\end{eqnarray}
 then $\sigma_1
\parallel \sigma_2$ is defined by cases as follows:

\begin{enumerate}
\item If both $\sigma_1$ and $\sigma_2$ have length $1$ and have
the  same store, say $\sigma_1 = \la c, \emptyset, H_1,c\ra$ and
$\sigma_2 = \la c, \emptyset, H_2,c\ra$, then
\[ \sigma_1 \parallel \sigma_2 =
\{\la c, \emptyset, H_1 \uplus H_2,c\ra\}.\]
\item If $\sigma_2$ has length $1$ and $\sigma_1$ has length $>1$ then
\[ \sigma_1 \parallel \sigma_2 =
\begin{array}[t]{l}
\{\la c_1, K_1, H_1 \uplus H_2,d_1\ra \cdot \sigma \in {\cal D}
\mid \ \sigma \in \sigma'_1 \parallel \sigma_2\}.
\end{array}
\]
The symmetric case is analogous and therefore omitted.\\
\item If both $\sigma_1$ and $\sigma_2$ have length $>1$ then
\[\begin{array}{lll}
\sigma _1\parallel \sigma_2 =& \{
 \la c_1, K_1,  H_1\uplus H_2, d_1\ra\cdot
 \sigma \in {\cal D} \mid  \sigma \in
\sigma'_1 \parallel \sigma_2\}
\\
&\cup
\\
& \{ \la c_2, K_2,  H_1\uplus H_2, d_2\ra\cdot \sigma \in {\cal D}
\mid \sigma \in \sigma_1 \parallel \sigma'_2\}
\end{array}
\]
\end{enumerate}

Finally the composition of sets of sequences $\parallel: \wp({\cal
D}) \times  \wp({\cal D}) \rightarrow \wp({\cal D})$ is defined by
\[\begin{array}{lll }
 S_1 \parallel S_2 =  \{\sigma \in {\cal D}\mid &
\mbox{there exist }\sigma_1\in S_1 \mbox{ and } \sigma_2\in S_2
\hbox{ such that } \\
& \sigma=\la c_1, K_1, H_1, d_1\ra  \cdots \la c_n, \emptyset,
H_n,c_n\ra \in \eta(\sigma_1 \parallel \sigma_2), \\
&  (V_{loc}(\sigma_1) \cup V_{loc}(\sigma_2)) \cap
V_{ass}(\sigma)=\emptyset  \mbox{ and for } i \in [1,n] \\
& (V_{loc}(\sigma_1) \cup V_{loc}(\sigma_2)) \cap Fv(c_i)
\subseteq \bigcup _{j=1} ^ {i-1} Fv(d_j) \cup Fv(H_i)\}.
\end{array}
\]
\end{definition}

Let us briefly illustrate some points in previous definition.

Condition (\ref{eqnuno}) ensures that the rules used to construct
the (derivations abstracted by the) sequences $\sigma_1$ and
$\sigma_2$ have been renamed apart (that is, they do not share
variables). Moreover, the local variables of each sequence are
different from those which appear in the initial goal for the
other sequence.

Moreover, in the definition of  the composition of sets of
sequences $\parallel: \wp({\cal D}) \times  \wp({\cal D})
\rightarrow \wp({\cal D})$, the first condition ensures that the
variables appearing in the rules used to construct the sequences
$\sigma_1$ and  $\sigma_2$ are distinct from the variables
appearing in the assumptions. The second condition is needed to
ensure that $\sigma$ is the abstraction of a sequence satisfying
condition 4 in Definition \ref{def:compatibility} (compatibility).

\bigskip

Using this notion of composition of sequences we can show that the
semantics ${\cal S}_P$ is compositional. Before proving the
compositionality theorem we need some technical lemmas.

\begin{lemma}\label{lem:stessevar}
Let $G$ be a goal, $\delta\in  {\cal S}'_P(G)$ and let $\sigma =
\alpha(\delta)$. Then $V_r(\delta)= V_r(\sigma)$ holds, where $r
\in \{\, ass, \, stable, \, constr, \, loc \, \}$.
\end{lemma}

\begin{lemma}\label{lem:percomp1}
Let $P$ be a program, $H$ and $G$ be two goals and assume that
$\delta\in  {\cal S}'_P(H,G)$. Then there exists $\delta_1\in
{\cal S}'_P(H)$ and $\delta_2\in  {\cal S}'_P(G)$, such that for
$i =1,2$, $V_{loc}(\delta_i) \subseteq V_{loc}(\delta)$ and
$\alpha(\delta) \in \eta (\alpha(\delta_1) \parallel
\alpha(\delta_2))$.
\end{lemma}

\begin{lemma}\label{lem:percomp2}
Let $P$ be a program, let $H$ and $G$ be two goals and assume that
$\delta_1\in  {\cal S}'_P(H)$ and $\delta_2\in {\cal S}'_P(G)$ are
two sequences such that the following hold:
\begin{enumerate}
    \item $\alpha(\delta_1) \parallel \alpha(\delta_2)$ is defined,
    \item $\sigma= \la c_1, K_1 ,W_1,d_1\ra \cdots
    \la c_n, \emptyset ,W_n,c_n\ra \in
    \eta (\alpha(\delta_1) \parallel \alpha(\delta_2))$,
    \item $(V_{loc}(\alpha(\delta_1)) \cup
    V_{loc}(\alpha(\delta_2)))
    \cap V_{ass}(\sigma)=\emptyset $,
    \item for $i \in [1,n]$,
    $(V_{loc}(\alpha(\delta_1)) \cup V_{loc}(\alpha(\delta_2)))
    \cap Fv(c_i) \subseteq \bigcup _{j=1} ^ {i-1} Fv(d_j) \cup Fv(W_i)$.
\end{enumerate}
Then there exists $\delta\in {\cal S}'_P(H,G)$ such that
$\sigma=\alpha(\delta)$.
\end{lemma}

By using the above results we can prove the following theorem.

\begin{theorem}\label{th:comp1}[Compositionality]
Let $P$ be a program and let $H$ and $G$ be two goals. Then
$$ {\cal S}_P(H,G) =
{\cal S}_P(H)\, \parallel \, {\cal S}_P(G).$$
\end{theorem}
{\bf Proof} We prove the two inclusions separately. \\

{\bf (}${\cal S}_P(H,G) \subseteq {\cal S}_P(H)\, \parallel \,
{\cal S}_P(G)${\bf )}. Let $\sigma \in {\cal S}_P(H,G)$. By
definition of ${\cal S}_P$, there exists $\delta \in {\cal
S}'_P(H,G)$ such that $\sigma =\alpha(\delta)$. By
Lemma~\ref{lem:percomp1} there exist $\delta_1 \in {\cal S}'_P(H)$
and $\delta_2 \in {\cal S}'_P(G)$ such that for $i =1,2$,
$V_{loc}(\delta_i) \subseteq V_{loc}(\delta)$ and $\sigma \in
\eta(\alpha(\delta_1) \parallel \alpha(\delta_2))$. Let
\[\delta= \la (H,G), c_1, K_1,B_2,d_1\ra \cdots
\la B_n, c_n, \emptyset,B_n, c_n\ra\] and let $\sigma=\la c_1,
K_1, H_1, d_1\ra \cdots \la c_n, \emptyset, H_n,c_n\ra$, where
$H_n=B_n$. Then in order to prove the thesis we have only to show
that
\[
\begin{array}{l}
(V_{loc}(\alpha(\delta_1)) \cup V_{loc}(\alpha(\delta_2))) \cap
V_{ass}(\sigma)=\emptyset  \mbox{ and  for $i \in [1,n]$, }\\
(V_{loc}(\alpha(\delta_1)) \cup V_{loc}(\alpha(\delta_2))) \cap
Fv(c_i) \subseteq \bigcup _{j=1} ^ {i-1} Fv(d_j) \cup  Fv(H_i).
\end{array}
\]
First observe that by Lemma~\ref{lem:stessevar} and by hypothesis,
we have that $V_{ass}(\sigma)=V_{ass}(\delta)$  and for $i=1,2$, $
V_{loc}(\alpha(\delta_i))=
    V_{loc}(\delta_i)\subseteq V_{loc}(\delta).$
Then by the previous results and by the properties of the
derivations
\[(V_{loc}(\alpha(\delta_1)) \cup V_{loc}(\alpha(\delta_2))) \cap
V_{ass}(\sigma)\subseteq V_{loc}(\delta) \cap
V_{ass}(\delta)=\emptyset.
\]
Moreover by condition 4 of Definition \ref{def:compatibility}
(compatibility), for $i \in [1,m]$,
\[(V_{loc}(\alpha(\delta_1)) \cup
V_{loc}(\alpha(\delta_2))) \cap Fv(c_i) \subseteq V_{loc}(\delta)
\cap Fv(c_i)\subseteq \bigcup _{j=1}^ {i-1} Fv(d_j) \cup
V_{stable}(\delta)
\]
holds. Now, observe that if $x \in V_{loc}(\delta) \cap Fv(c_i)
\cap V_{stable}(\delta)$, then $x \in \bigcup _{j=1} ^ {i}
V_{loc}(\delta) \cap Fv(B_j)\cap V_{stable}(\delta)$ and then $x
\in Fv(H_i)$ and this
completes the proof of the first inclusion.\\

{\bf (}$ {\cal S}_P(H,G) \supseteq  {\cal S}_P(H)\,
\parallel \, {\cal S}_P(G)${\bf )}. Let $\sigma \in {\cal S}_P(H)\,
\parallel \, {\cal S}_P(G)$. By definition of ${\cal S}_P$ and
of $\parallel$ there exist $\delta_1 \in {\cal S}'_P(H)$ and
$\delta_2 \in {\cal S}'_P(G)$, such that
$\sigma_1=\alpha(\delta_1)$, $\sigma_2=\alpha(\delta_2)$,
$\sigma_1
\parallel \sigma_2$ is defined,
$\sigma=\la c_1, K_1, H_1, d_1\ra  \cdots \la c_n, \emptyset,
H_n,c_n\ra \in \eta(\sigma_1 \parallel \sigma_2)$,
$(V_{loc}(\sigma_1) \cup V_{loc}(\sigma_2)) \cap
V_{ass}(\sigma)=\emptyset$ and for $i \in [1,n]$,
$(V_{loc}(\sigma_1) \cup V_{loc}(\sigma_2)) \cap Fv(c_i) \subseteq
\bigcup _{j=1} ^ {i-1} Fv(d_j) \cup  Fv(H_i)$. The proof is then
straightforward by using Lemma~\ref{lem:percomp2}.

\subsection{Correctness}\label{subsec:correct}

In order to show the correctness of the semantics ${\cal S}_P$
w.r.t. the (input/output) observables ${\cal SA}_P$, we first
introduce a different characterization of ${\cal SA}_P$ obtained
by using the new transition system defined in Table \ref{tcs}.

\begin{definition}
Let $P$ be a program and let $G$ be a goal and let $\rrarrow_P$ be
(the least relation) defined by the rules in Table \ref{tcs}. We
define
\[{\cal SA'}_P(G) = \{ \exists_{-Fv(G)} c  \mid
\la G, \emptyset\ra \rrarrow_P ^\emptyset \cdots \rrarrow_P
^\emptyset \la \emptyset, c\ra \not \rrarrow_P^K \}.
\]
\end{definition}

The correspondence of ${\cal SA'}$ with the original notion ${\cal
SA}$ is stated by the following proposition, whose proof is
immediate.
\begin{proposition}\label{prop:corr}
Let $P$ be a program and let $  G$ be a goal. Then
\[{\cal SA}_P(G) = {\cal SA}'_P(G).
\]
\end{proposition}
The observables ${\cal SA}'_P$, and therefore ${\cal SA}_P$,
describing answers of ``data sufficient'' computations can be
obtained from ${\cal S}$ by considering suitable sequences, namely
those sequences which do not perform assumptions neither on CHR
constraints nor on built-in constraints. The first condition means
that the second components of tuples must be empty, while the
second one means that the assumed constraint at step i must be
equal to the produced constraint at step i-1. We call
``connected'' those sequences which satisfy these requirements:

\begin{definition}\label{def:connected}[Connected sequences]
Assume that
\[\sigma = \la c_1,K_1,H_1,d_1\ra \ldots \la
c_n,K_n,H_n,c_n\ra \] is a sequence in ${\cal D}$. We say that
$\sigma$ is connected if
\begin{enumerate}
    \item $K_i=\emptyset$ for each $i$, $1\leq i\leq n$,
    \item $d_j = c_{j+1}$ for each $j$, $1\leq j\leq n-1$ and
    \item either $H_n=\emptyset$ or $c_n = {\tt false}$.
\end{enumerate}
\end{definition}

The proof of the following result derives from the definition of
connected sequence and an easy inductive argument. \\
Given a sequence $\sigma = \la c_1,K_1,H_1,d_1\ra \ldots \la
c_n,K_n,H_n,d_n\ra$, we denote by $instore(\sigma)$ and
$store(\sigma)$ the built-in constraint $c_1$ and the built-in
constraint $d_n$, respectively.

\begin{proposition}
Let $P$ be a program and let $ G$ be a goal. Then
\[ {\cal SA}'_P(G) =\{\exists_{-Fv(G)} c  \mid
\begin{array}[t]{l} \mbox{there exists } \sigma
\in {\cal S}_P(G) \mbox{ such that } instore(\sigma)=\emptyset \\
\sigma \mbox{ is connected and } c=store(\sigma) \}.
\end{array}
\]
\end{proposition}

The following corollary is immediate from
Proposition~\ref{prop:corr}.

\begin{corollary}\label{cor:correct}[Correctness]
Let $P$ be a program and let $ G$ be a goal. Then
\[ {\cal SA}_P(G) =\{\exists_{-Fv(G)} c  \mid
\begin{array}[t]{l} \mbox{there exists }
\sigma \in {\cal S}_P(G) \mbox{ such that }
instore(\sigma)=\emptyset \\
\sigma \mbox{ is connected and } c=store(\sigma) \}.
\end{array}
\]
\end{corollary}

\section{A more refined semantics}\label{sec:refined}

As previously mentioned, the operational semantics that we have
considered in this paper is somehow naive: In fact, since
propagation rules do not remove user defined constraints (see rule
Propagate in Table \ref{tor}), when a propagate rule is applied it
introduces an additional infinite  computation (obtained by
subsequent applications of the same rule). Of course, as
previously mentioned, the terminating computations are not
affected, as  the application of a simplification rule after a
propagation rule can cause the termination of the computation.

A more refined operational
semantics which avoid these infinite computations has been defined in
 \cite{Ab97}.
Essentially the idea is to memorize in a {\em token store}, to be
added to the global state,  some {\em tokens} containing the
information about which propagation rules can be applied to a
given multiset of user-defined constraints. Each {\em token}
consists of a propagation rule name and of the multiset of
candidate constraints for that rule. A propagation rule can then
be applied only if the store contains the appropriate token and therefore it
can be applied at most once to the same constraint.

We could take into account this refined operational semantics by
using a slight extension of our semantic construction. More
precisely, we first consider ``concrete'' sequences consisting of
tuples of the form $\la G,c,T,K,G',T',d\ra$, where $T$ and $T'$
are token stores as defined in \cite{Ab97}. Such a tuple
represents exactly a derivation step $\la G,c, T\ra \rrarrow_P^K
\la G',d, T'\ra$, according to the operational semantics in
\cite{Ab97}. The sequences we consider are terminated by tuples of
the form $\la G,c,T,\emptyset,G,c, T\ra$, which represent a
terminating step. Since a sequence represents a derivation, we
assume that the ``output'' goal $G'$ and token store $T'$ at step
$i$ are equal to the ``input'' goal $G$ and to the token store $T$
at step $i+1$, respectively. From these concrete sequences we
extract the same abstract sequences which are the objects of our
semantic domain: From each tuple $\la G,c,T,K,G',d, T'\ra$ in a
concrete sequence $\delta$ we extract  a tuple of the form $\la
c,K,T,H,d\ra$ where we consider as before the input and output
store ($c$ and $d$, respectively), the input token store and the
assumptions ($K$), while we do not consider anymore the output
goal $G'$ and the token store $T'$. The abstraction operator which
extracts from the concrete sequences the sequences used in the
semantic domain is a simple extension to that one given in
Definition~\ref{def:alpha}. In order to obtain a compositionality
result we then define how two sequences describing a computation
of $A$ and $B$ according to this refined operational semantics,
respectively, can be composed in order to obtain a computation of
$A,B$. Such a composition is defined by a (semantic) operator,
which performs an interleaving of the actions described by the two
sequences. This new operator is similar to that one defined in
Definition~\ref{def:composition} even though the technicalities
are different.

Recently a more refined semantics has been defined in \cite{DGS04}
in order to describe precisely the operational semantics
implicitly used by (Prolog) implementations of CHR. Although this
refined operational semantics is still non-deterministic, the
order in which transitions are applied and the order in which
occurrences are visited are decided. This semantics is therefore
substantially different from the one we consider and apparently it
is difficult to give a compositional characterization for it.

\section{Conclusions}\label{sec:concl}

In this paper we have introduced a semantics for CHR which is
compositional w.r.t. the and-composition of goals and which is
correct w.r.t ``data sufficient answers'', a notion of observable
which considers the results of (finitely) failed computations and
of successful computations where all the user-defined constraints
have been rewritten into built-in constraints. We are not aware of
other compositional characterizations of CHR answers and only
\cite{Ma02} addresses compositionality of CHR rules (but only for
a subset of CHR). Our work can be considered as a first step which
can be extended along several different lines.

Firstly, it would be desirable to obtain a compositional
characterization also for ``qualified answers'' obtained by
considering computations terminating with a user-defined
constraint which does not need to be empty (see
Definition~\ref{opsemqa}). This could be done by a slight
extension of our model: The problem here is that, given a tuple
$\la G,c,K,G',d\ra$, in order to reconstruct correctly the
qualified answers we need to know whether the configuration $\la
G',d\ra$ is terminating or not (that is, if $\la G',d\ra
\not\rightarrow_P^{K'}$ holds). This could be solved by
introducing some termination modes, at the price of a further
complication of the traces used in our semantics. Also, as
previously mentioned, we are currently extending our semantics in
order do describe the more refined operational semantics given in
\cite{Ab97}.

A second possible extension is the investigation of the full
abstraction issue. For obvious reasons it would be desirable to
introduce in the semantics the minimum amount of information
needed to obtain compositionality, while preserving correctness.
In other terms, one would like to obtain a results of this kind:
${\cal S}_P(G) = {\cal S}_P(G')$ if and only if, for any $H$,
${\cal SA}_P(G,H) = {\cal SA}_P(G',H)$ (our Corollary
\ref{cor:correct} only ensures that the ``only if'' part holds).
Such a full abstraction result could be difficult to achieve,
however techniques similar to those used in \cite{BP90c,BGM97} for
analogous results in the context of ccp could be considered

It would be interesting also to study further notions of
compositionality, for example that one which considers union of
program rules rather than conjunctions of goals, analogously to
what has been done in \cite{BGLM92a}. However, due to the presence
of synchronization, the simple model based on clauses defined in
\cite{BGLM92a} cannot be used for CHR.

As mentioned in the introduction, the main interest related to a
compositional semantics is the possibility to provide a basis to
define compositional analysis and verification tools. In our case,
it would be interesting to investigate to what extent the
compositional proof systems \emph{\`a la} Hoare defined in
\cite{BGMP94,BGM04} for timed ccp languages, based on resting
points and trace semantics, can be adapted to the case of CHR.
Also, it would be interesting to apply the semantics to
reconstruct the confluence analysis of CHR.

\noindent
 {\bf Acknowledgments} We thank Michael Maher for having
 initially suggested the problem of compositionality for CHR
 semantics.

\newpage

\section{Appendix}\label{sec:app}

In this appendix we provide the proofs of some lemmas used in the
paper.

In the following, given a sequence $\gamma$, where  $\gamma \in
{\cal S}eq \cup {\cal D}$, we will denote by $instore(\gamma)$ and
by $Inc(\gamma)$  the first input constraint and the set of input
constraints of $\gamma$, respectively. Moreover, we will denote by
$Ass(\gamma)$ and $Stable(\gamma)$ the set (corresponding to the
multiset) of assumptions of $\gamma$ and the set (corresponding to
the multiset) of atoms in the last goal of $\gamma$, respectively.

\begin{lemma} (Lemma \ref{lem:stessevar})
Let $G$ be a goal, $\delta\in  {\cal S}'_P(G)$ and let $\sigma =
\alpha(\delta)$. Then \[ V_r(\delta)= V_r(\sigma) \mbox{, where }
r \in \{ \, ass, \, stable, \, constr, \, loc \, \}.\]
\end{lemma}
{\bf Proof} If $r \in \{ \, ass, \, stable, \, constr \, \}$ then
the proof is straightforward by definition of $\alpha$ and of
$V_r$. Then we have only to prove that $V_{loc}(\delta)=
V_{loc}(\sigma)$. \\
The proof is by induction on $n=lenght(\delta)$.
\begin{description}
    \item[$n=1$)] In this case $\delta = \la G,c, \emptyset,
    G,c\ra$, $\sigma = \la c, \emptyset,
    G,c\ra$, and therefore, by definition $V_{loc}(\delta)=
    V_{loc}(\sigma) = \emptyset.$
    \item[$n\geq 1$)] Let $\delta = \la G_1,c_1, K_1, G_2,d_1\ra
    \la G_2,c_2, K_2, G_3,d_3\ra \cdots \la G_n, c_n,\emptyset,G_n,c_n\ra$,
    where $G=G_1$. \\
    By definition of $ {\cal S}'_P(G)$, there exists
    $\delta'\in  {\cal S}'_P(G_2)$
    such that $t=\la G_1,c_1, K_1, G_2,d_1\ra$ is compatible with
    $\delta'$ and $ \delta = t \cdot
    \delta' \in {\cal S}eq$.

    By inductive hypothesis, we have that
    $V_{loc}(\delta')= V_{loc}(\sigma')$, where
    $\sigma'=\alpha(\delta')$. \\
    Moreover, by definition of
    $\alpha$,
    $\sigma = \la c_1,K_1,H_1,d_1\ra\cdot \sigma'$,
    where $H_1$ is the multiset consisting of all the atoms in $G_1$
    which are stable in $\delta$.

    By definition of $V_{loc}$ and by inductive hypothesis
    \begin{eqnarray}
    \nonumber
    V_{loc}(\delta) & =& \bigcup_{i=1}^{n-1} Fv(G_{i+1}, d_i)
    \setminus Fv(G_i,c_i, K_i)\\
    \nonumber  & =&
    V_{loc}(\delta') \cup (Fv(G_2, d_1) \setminus Fv(G_1,c_1, K_1)) \\
    \label{eq:21feb1}
    & =&  V_{loc}(\sigma') \cup (Fv(G_2, d_1) \setminus Fv(G_1,c_1, K_1)).
    \end{eqnarray}

    Moreover, by definition of $V_{loc}$ and since
    $V_{stable}(\sigma)=V_{stable}(\sigma')$,
    we have that
    \begin{eqnarray}
    \label{eq:3maggio1}
    V_{loc}(\sigma')=(V_{constr}(\sigma')\cup
    V_{stable}(\sigma))
    \setminus(V_{ass}(\sigma')\cup Fv(G_2)).
    \end{eqnarray}
    Therefore by (\ref{eq:21feb1}),
    by properties of $\cup$ and since $Fv(G_2)\cap Fv(G_1,c_1,
    K_1)\subseteq Fv(G_2)\cap Fv(G_1)$, we have that
    \begin{eqnarray}
     \nonumber
       V_{loc}(\delta) & = &
     ((V_{constr}(\sigma')\cup
      V_{stable}(\sigma))
      \setminus(V_{ass}(\sigma')\cup Fv(G_2)))\ \cup \\
      \label{eq:21feb22}
        & & (Fv(G_2) \setminus Fv(G_1)) \  \cup \
        (Fv(d_1) \setminus Fv(G_1,c_1,K_1)).
    \end{eqnarray}
    Now, let $x \in Fv(K_1) \cap (V_{constr}(\sigma')\cup
    V_{stable}(\sigma))$. By definition
    $x \in Fv(t)$, since $t$ is compatible with $\delta'$ and by
    condition 2 of Definition \ref{def:compatibility} (compatibility), we have that
    $x \not \in V_{loc}(\delta')=V_{loc}(\sigma')$ and therefore
    by (\ref{eq:3maggio1}) $x \in V_{ass}(\sigma')\cup Fv(G_2)$.
    Then by (\ref{eq:21feb22})
    \begin{eqnarray}
     \nonumber
      V_{loc}(\delta) &=& ((V_{constr}(\sigma')\cup V_{stable}(\sigma))
      \setminus(V_{ass}(\sigma)\cup Fv(G_2)))\\
      \label{eq:21feb2}
        &  & \cup \ (Fv(G_2) \setminus Fv(G_1,K_1)) \  \cup \ (Fv(d_1) \setminus Fv(G_1,c_1,
        K_1)).
    \end{eqnarray}
    By properties of $\cup$, we have that
    \begin{eqnarray}
      \nonumber & &((V_{constr}(\sigma')\cup
      V_{stable}(\sigma))
      \setminus V_{ass}(\sigma)\cup Fv(G_2)))  \cup
       \\
      \nonumber & & (Fv(G_2) \setminus
      Fv(G_1))  \ = \\
      \nonumber && (( V_{constr}(\sigma')\cup
      V_{stable}(\sigma))
      \setminus (V_{ass}(\sigma)\cup (Fv(G_2)\cap Fv(G_1))))
      \cup \\
      & &   (Fv(G_2) \setminus
      Fv(G_1)). \label{eq:21feb3}
      \end{eqnarray}
    Now let $x \in Fv(G_1) \setminus \, Fv(G_2)$ and let us assume that $x \in
      V_{constr}(\sigma') \cup V_{stable}(\sigma)
      =V_{constr}(\delta')\cup V_{stable}(\delta')$.
    By definition
    $ x \in Fv(t)$, since $t$ is compatible with $\delta'$ and by
    condition 2 of Definition \ref{def:compatibility} (compatibility), we have that
    $x \not \in V_{loc}(\delta')$. Then since $x \not \in Fv(G_2)$
    we have that there exists $i \in [2,n-1]$
    such that $x \in Fv(K_i)$ and therefore $x \in
    V_{ass}(\delta')= V_{ass}(\sigma')$.
    Therefore, by the previous results and by (\ref{eq:21feb2}) and
      (\ref{eq:21feb3}), we have that
      \begin{eqnarray}
     \nonumber
     V_{loc}(\delta) &= &
     ((V_{constr}(\sigma')\cup
     V_{stable}(\sigma))
      \setminus(V_{ass}(\sigma)\cup Fv(G_1))) \cup \ \\
      \label{eq:21feb4}
        &  &(Fv(G_2) \setminus Fv(G_1)) \  \cup \
        (Fv(d_1) \setminus Fv(G_1,c_1,  K_1)).
    \end{eqnarray}
    Now let $x \in (Fv(d_1) \setminus Fv(c_1)) \cap  V_{ass}(\sigma')$.
    Since by point 3 of Definition \ref{def:compatibility}
    (ompatibility)
    $V_{loc}( t )\cap V_{ass}(\sigma') =
    \emptyset$, we have that $ x \in Fv(G_1,K_1)$. Then
    \[\begin{array}{lll}
      Fv(d_1) \, \setminus\, Fv(G_1, c_1, K_1) & = & \\
     (Fv(d_1)\, \setminus \, Fv(c_1)) \, \setminus \, Fv(G_1,
     K_1)\
       & = & \\
       (Fv(d_1)\, \setminus \, Fv(c_1)) \, \setminus \,
       (Fv(G_1, K_1) \cup  V_{ass}(\sigma'))\
       & = & \\
       (Fv(d_1)\, \setminus \, Fv(c_1)) \, \setminus \, (Fv(G_1) \cup
       V_{ass}(\sigma)).
    \end{array}
     \]
     Then by (\ref{eq:21feb4}),
     \begin{eqnarray}
     \nonumber
     V_{loc}(\delta)  &=&
     ((V_{constr}(\sigma)\cup V_{stable}(\sigma)) \setminus
     (V_{ass}(\sigma)\cup Fv(G_1))) \ \cup \\
     && (Fv(G_2) \setminus Fv(G_1)).
     \label{eq:21feb6}
     \end{eqnarray}
     Finally let $x \in Fv(G_2) \setminus Fv(G_1)$.
     We prove that $x \in ((V_{constr}(\sigma)\cup V_{stable}(\sigma)) \setminus
     V_{ass}(\sigma)$. First of all, observe that $x \in V_{loc}(t)$ and therefore,
     by definition of compatibility, $x \not \in V_{ass}(\sigma)$.
     Now, let $A \in G_2$ such that $x \in Fv(A)$
     and let us to assume that $A \not \in
     Stable(\sigma)= Stable(\delta)$. Then, by definition of derivation,
     there exists $j \in [1,n-1]$ such that $x \in
     Fv(d_j)$. Let $h$ the least index $j \in [1,n-1]$ such that
     $x \in Fv(d_h)$. By condition 4 of Definition \ref{def:compatibility} (compatibility),
     we have that $x \not\in Fv(c_h)$ and then $x \in
     V_{constr}(\delta)=V_{constr}(\sigma)$.
     Then by (\ref{eq:21feb6}), by the previous result
     and by definition of $V_{loc}$,
     \[\begin{array}{ll}
       V_{loc}(\delta)  =
      (V_{constr}(\sigma)\cup V_{stable}(\sigma))
      \setminus(V_{ass}(\sigma)\cup Fv(G_1))  =
      V_{loc}(\sigma)
     \end{array}
      \]
      and then the thesis holds.
\end{description}

\bigskip

In the following,  given a sequence of derivation steps
\[\delta = \la B_1, c_1, K_1,B_2,d_1\ra \ldots \la B_n, c_n,
\emptyset,B_n,c_n\ra\]
and a goal $W$, we denote by $\delta\oplus
W$ the sequence
\[\la (B_1,W), c_1, K_1,(B_2,W),d_1\ra
\ldots \la (B_n,W), c_n, \emptyset,(B_n,W),c_n\ra \] and by
$\delta\ominus W$ the sequence
\[\la B_1\setminus W, c_1, K_1,B_2\setminus W,d_1\ra
\ldots \la B_n\setminus W, c_n, \emptyset,B_n\setminus W,c_n\ra .\]

The proof of the following two lemma is straightforward by
definition of derivation.

\begin{lemma}\label{lem:dagoalasass}
Let $H,G$ be goals and let $\delta\in  {\cal S}'_P(H,G)$ such that
\[\begin{array}{lll}
\delta & = & \la (H,G), c_1, K_1,R_2, d_1 \ra\la R_2, c_2,K_2,R_3,
d_2\ra \cdots \la R_n, c_n, \emptyset, R_n, c_n \ra
\end{array}  \]
where $H=(H', H'')$, $H''\neq \emptyset$ and the first tuple of
the sequence $\delta$ represents a derivation step $s$, which uses
the {\bf Apply'} rule and rewrites only and all the atoms in
$(H'',G)$. Then there exists a derivation $\delta '\in {\cal
S}'_P(H)$ such that
\[\begin{array}{lll}
\delta' & = & \la H, c_1,K_1 \uplus G, R_2, d_1\ra\la R_2,
c_2,K_2,R_3, d_2\ra  \cdots \la R_n, c_n \emptyset,R_n, c_n \ra.
\end{array} \]
\end{lemma}

\begin{lemma}\label{lem:aggoal}
Let $G$ be a goal, $W$ be a multiset of atoms and let $\delta\in
{\cal S}'_P(G)$ such that $Fv(W) \cap V_{loc}(\delta)=\emptyset$.
Then $\delta\oplus W \in {\cal S}'_P(G, W)$.
\end{lemma}

\begin{lemma} \label{lem:ingoalH}
Let $P$ be a program and let $H$ and $G$ be two goals such that
there exists a derivation step
\[ s=\la (H,G), c_1 \ra
\rrarrow_P^{K_1} \la (B, G),d_1\ra,\]
where only the atoms in $H$ are rewritten in $s$. \\
Assume that there exists $\delta \in {\cal S}'_P(H,G)$ such that
$\delta=t \cdot \delta'$, where
\[t=\la (H,G), c_1,
K_1, (B, G),d_1 \ra, \] $\delta' \in {\cal S}'_P(B,G)$ and $t$ is
compatible with $\delta'$. Moreover assume that there exists
$\delta'_1\in {\cal S}'_P(B)$ and $\delta'_2\in {\cal S}'_P(G)$,
such that
    \begin{enumerate}
    \item for $i=1,2$, $V_{loc}(\delta'_i)\subseteq
    V_{loc}(\delta')$
    and $Inc(\delta'_i)\subseteq Inc(\delta')$.
    \item $Ass(\delta'_1)\subseteq Ass(\delta') \cup Stable (\delta'_2)$
    and $Ass(\delta'_2)\subseteq Ass(\delta') \cup Stable (\delta'_1)$,
    \item
    $\alpha(\delta'_1) \parallel \alpha(\delta'_2)$ is defined and
    $\alpha(\delta') \in \eta (\alpha(\delta'_1) \parallel
\alpha(\delta'_2))$.
\end{enumerate}
Then $\delta_1=t'\cdot \delta'_1 \in {\cal S}'_P(H)$, where
$t'=\la H, c_1, K_1, B,d_1 \ra,$ $\alpha(\delta_1)
\parallel \alpha(\delta'_2)$ is defined and
$\alpha(\delta) \in \eta (\alpha(\delta_1) \parallel
\alpha(\delta'_2))$.
\end{lemma}
{\bf Proof } In the following, assume that
  \[\begin{array}{l}
    \delta'_1= \la B_1, e_1,M_1, B_2,f_1\ra \la B_2, e_2,M_2, B_3,f_2\ra \cdots \la
    B_l, e_l,\emptyset, B_l, e_l,\ra\\
    \delta'_2= \la G_1, r_1,N_1, G_2,s_1 \ra \la G_2, r_2,N_2, G_3,s_2\ra \cdots \la G_p, r_p,
    \emptyset, G_p, r_p\ra\\
    \delta'= \la  R_1, c_2,K_2, R_2,d_2 \ra \la  R_2, c_3,K_3, R_3,d_3 \ra \cdots
    \la R_{n-1}, c_n,\emptyset, R_{n-1}, c_n\ra, \\
  \end{array}
  \]
  where $B_1=B$, $G_1=G$, $R_1=(B,G)$ and $e_l=r_p=c_n$. The following holds.
\begin{description}
    \item[{\bf (a)} $\delta_1 \in {\cal S}'_P(H)$]
   By construction, we have only to prove that $t'$ is compatible
   with $\delta'_1$. The following holds.
   \begin{enumerate}
    \item By hypothesis $Inc(\delta'_1)\subseteq Inc(\delta')$ and
then $CT \models instore (\delta'_1)  \rightarrow instore
(\delta')$.  Moreover since $t$ is compatible with $\delta'$, we
have that $CT \models instore (\delta')  \rightarrow d_1$ and
therefore $CT \models instore (\delta'_1)  \rightarrow d_1$.
    \item By hypothesis $V_{loc}(\delta'_1)\subseteq
V_{loc}(\delta')$ and by construction $Fv(t')\subseteq Fv(t)$.
Then $V_{loc}(\delta'_1) \cap Fv(t')\subseteq V_{loc}(\delta')
\cap Fv(t) = \emptyset$, where the last equality follows since $t$
is compatible with $\delta'$.

    \item First of all observe that given a derivation $\tilde\delta$, we have that
\begin{equation}\label{eq:28aprile2}
    V_{Stable} (\tilde\delta) \subseteq Fv(\tilde G) \cup V_{loc}(\tilde \delta),
\end{equation}
where $\tilde G$ is the initial goal of the derivation
$\tilde\delta$. Then have that
         \[\begin{array}{lll}
         V_{loc}(t') \cap V_{ass}(\delta'_1) \subseteq & \\
         \hspace{0.5cm} \mbox{(since  $V_{loc}(t')=V_{loc}(t)$
         and since by hypothesis}\\
         \hspace{0.5cm} \  Ass(\delta'_1)\subseteq Ass(\delta') \cup Stable(\delta'_2))\\
         V_{loc}(t) \cap (V_{ass}(\delta') \cup V_{Stable}
         (\delta'_2)) \subseteq &\\
         \hspace{0.5cm} \mbox{(by (\ref{eq:28aprile2}))} \\
         V_{loc}(t) \cap (V_{ass}(\delta') \cup Fv(G)
         \cup V_{loc}(\delta'_2)) \subseteq &\\
         \hspace{0.5cm}  \mbox{(since by hypothesis
         $ V_{loc}(\delta'_2)\subseteq V_{loc}(\delta')$)}\\
         V_{loc}(t) \cap (V_{ass}(\delta') \cup Fv(G)
         \cup V_{loc}(\delta')) = &\\
         \hspace{0.5cm} \mbox{(since $t$ is compatible with $\delta'$ and by definition of
         $V_{loc}$)}\\
         \emptyset
         \end{array}
         \]

\item  We have to prove that for $i \in [1,l]$, $V_{loc}(t') \cap
Fv(e_i)\subseteq \bigcup _{j=1}^{i-1} Fv(f_j) \cup Fv(d_1) \cup
V_{stable}(\delta'_1)$. Let $i \in [1,l]$ and let $x \in
V_{loc}(t') \cap Fv(e_i)$.

Since by inductive hypothesis $Inc (\delta'_1) \subseteq
Inc(\delta')$, there exists a least index $h \in [2,n]$ such that
$e_i=c_h$. Therefore, since $V_{loc}(t')=V_{loc}(t)$ and $t$ is
compatible with $\delta'$, we have that
\begin{equation}\label{eq:28aprile1}
    x \in \bigcup _{j=1}^{h-1} Fv(d_j) \cup V_{stable}(\delta').
\end{equation}
Moreover, since $x \in V_{loc}(t')=V_{loc}(t)$, $t$ is compatible
with $\delta'$ and by hypothesis $V_{loc}(\delta'_2) \subseteq
V_{loc}(\delta')$
\begin{equation}\label{eq:1maggio31}
    x \not \in Fv(G) \cup V_{loc}(\delta'_2).
\end{equation}
Now, observe that
\[\begin{array}{lll}
   V_{stable}(\delta') &\subseteq &
   \mbox{(by definition of
  $\parallel$ and since by hypothesis }\\
  && \mbox{
  $\alpha(\delta')\in \eta(\alpha(\delta'_1) \parallel
    \alpha(\delta'_2))$)} \\
   V_{stable}(\delta'_1)\cup V_{stable}(\delta'_2) & \subseteq  &
   \mbox{(by (\ref{eq:28aprile2}))}\\
  V_{stable}(\delta'_1)\cup Fv(G
  ) \cup V_{loc}(\delta'_2).
\end{array}
\]
Then by (\ref{eq:28aprile1}) and (\ref{eq:1maggio31}), we have
that $ x \in \bigcup _{j=1}^{h-1} Fv(d_j) \cup
V_{stable}(\delta'_1)$.
Then to prove the thesis, we have to prove that \\
if $x \in \bigcup _{j=1}^{h-1} Fv(d_j) \cup V_{stable}(\delta'_1)$
then $x \in \bigcup _{j=1}^{i-1} Fv(f_j)\cup Fv(d_1) \cup
V_{stable}(\delta'_1)$.

Let us to assume that $x \in \bigcup _{j=2}^{h-1} Fv(d_j)$  and
let $k$ the least index $j \in [2,h-1]$ such that $x \in Fv(d_j)$.

If $d_k$ is an output constraint of $\delta'_1$, i.e. there exists
$j \in [1,i-1]$ such that $d_k=f_j$, the proof is terminated.

Now assume that $d_k$ is an output constraint of $\delta'_2$, i.e.
there exists $w \in [1,m]$ such that $d_k=s_w$ and for each $j \in
[1,w-1]$, we have that $x \not \in Fv(s_j)$. Since $k$ is the
least index $j$ such that $x \in Fv(d_j)$ and since $t$ is
compatible with $\delta'$, we have that $x \not \in Fv(c_k)$ and
therefore $x \not \in Fv(r_w)$.

Moreover, since by (\ref{eq:1maggio31}), $x \not \in Fv(G) \cup
V_{loc}(\delta'_2)$, we have that $x \not \in Fv(G_w)$. Then by
definition of derivation step, since $x \in Fv(s_w) \setminus
(Fv(r_w) \cup Fv(G_w))$, we have that $x \in Fv(N_w)$ and
therefore $x \in V_{ass}(\delta'_2)$. By hypothesis $x \in
V_{ass}(\delta') \cup V_{stable}(\delta'_1)$. Then since $t$ is
compatible with $\delta'$ and $x \in V_{loc}(t)$, we have that $x
\not \in V_{ass}(\delta')$ and therefore  $x \in
V_{stable}(\delta'_1)$ and then the proof.
\end{enumerate}

    \item[{\bf (b) } $\alpha(\delta_1)
\parallel \alpha(\delta'_2)$ is defined] We have to prove that
\[
(V_{loc}(\alpha(\delta_1)) \cup Fv(H)) \cap
(V_{loc}(\alpha(\delta_2')) \cup Fv(G))\subseteq Fv(H) \cap Fv(G).
\]
By Lemma~\ref{lem:stessevar}
\begin{equation}\label{eq:1maggio3}
   V_{loc}(\alpha(\delta_1)) =  V_{loc}(\alpha(\delta'_1)) \cup  V_{loc}(t')
\end{equation} and
since $\alpha(\delta'_1)
\parallel \alpha(\delta'_2)$ is defined , we have that
\begin{equation}\label{eq:1maggio2}
   V_{loc}(\alpha(\delta'_1))  \cap
   (V_{loc}(\alpha(\delta'_2)) \cup Fv(G))=\emptyset .
\end{equation}
Now observe that, since $t$ is compatible with $\delta'$,
$V_{loc}(t')=V_{loc}(t)$ and by Lemma~\ref{lem:stessevar}, we have
that $V_{loc}(t') \cap V_{loc}(\alpha(\delta'))=\emptyset$.
Moreover, by hypothesis for $V_{loc}(\alpha(\delta'_2)) \subseteq
V_{loc}(\alpha(\delta'))$ and by definition of $t$, we have that
$Fv(G)\cap V_{loc}(t')=Fv(G)\cap V_{loc}(t)=\emptyset$. Then
\[
\begin{array}{l} V_{loc}(\alpha(\delta_1)) \cap (V_{loc}(\alpha(\delta'_2))
\cup Fv(G))= \\
( V_{loc}(\alpha(\delta'_1)) \cup V_{loc}(t')) \cap
(V_{loc}(\alpha(\delta'_2)) \cup Fv(G)) =\emptyset.
\end{array}
\]
Moreover, since $t$ is compatible with $\delta'$, $Fv(H) \subseteq
Fv(t)$ and by hypothesis $V_{loc}(\alpha(\delta'_2)) \subseteq
V_{loc}(\alpha(\delta'))$
\[Fv(H) \cap
V_{loc}(\alpha(\delta'_2)) \subseteq Fv(H) \cap
V_{loc}(\alpha(\delta')) = \emptyset
\]
and then the thesis holds.

    \item[{\bf (c)} $ \alpha(\delta) \in
         \eta(\alpha(\delta_1) \parallel
         \alpha(\delta'_2))$]
         By hypothesis $ \alpha(\delta') \in \eta(\alpha(\delta_1') \parallel \alpha(\delta'_2))$,
         $\alpha(\delta)= \la c_1, K_1,W_1,d_1\ra \cdot
         \alpha(\delta')$ and
         $\alpha(\delta_1)=
         \la c_1, K_1,J_1,d_1\ra \cdot
         \alpha(\delta'_1)$, where  $W_1$ is is the multiset of
         atoms in $(H,G)$ which are not rewritten in
         $\delta$ and $J_1$ is the multiset of
         atoms in $H$ which are not rewritten in
         $\delta_1$. Moreover let us to denote by
         \begin{itemize}
            \item  $J_2$ the set of atoms in $B$ which are
            not rewritten in $\delta'_1$, by
            \item $Y_1$ the set of atoms in $G$ which are
            not rewritten in $\delta'_2$ and by
            \item $W_2$ the set of atoms in $(B,G)$ which are
            not rewritten in $\delta'$.
         \end{itemize}
         Since $\alpha(\delta')\in \eta(\alpha(\delta'_1) \parallel
         \alpha(\delta'_2))$ there exists
         $\sigma'\in {\cal D}$ such that
         \[
         \sigma'\in \alpha(\delta'_1) \parallel
         \alpha(\delta'_2)
         \mbox{  and }\
         \alpha(\delta') \in \eta(\{\sigma'\}).
         \]
         By our assumptions,
         $\sigma'= \la c_2, A_1, J_2\uplus Y_1, d_2 \ra \cdot \sigma''$
         and by definition of $\parallel$,
         \[\sigma= \la c_1, K_1,J_1 \uplus Y_1, d_1 \ra \cdot  \sigma' \in
         \alpha(\delta_1) \parallel \alpha(\delta'_2).
         \]
         By definition of $\eta$ and since $ \alpha(\delta') \in \eta(\{\sigma'\})$,
         \begin{eqnarray}
         \label{14luglio1}
            \la c_1, K_1,(J_1 \uplus Y_1) \setminus S, d_1 \ra \cdot  \alpha(\delta')\in
         \eta(\alpha(\delta_1) \parallel \alpha(\delta'_2)),
         \end{eqnarray}
         where the multisets difference $(J_1 \uplus Y_1) \setminus S$
         considers indexes and $S$ is such that
         $(J_2 \uplus Y_1) \setminus S= W_2$. Then we can choose
         $S$ in such a way that $S$ restricted to the atoms with
         index equal to $1$ is the set of (non-indexed) atoms $(J_1 \uplus Y_1) \setminus
         W_1$ and $S$ restricted to the atoms with
         index equal to $2$ is the set of (non-indexed) atoms $(J_2 \setminus J_1) \setminus
         (W_2 \setminus W_1)$. It is easy to check that $S$
         satisfies the condition $(J_2 \uplus Y_1) \setminus S=
         W_2$. Moreover, by construction $(J_1 \uplus Y_1) \setminus
         S=W_1$.
         Therefore by (\ref{14luglio1})
         \begin{eqnarray}
          \nonumber
           \alpha(\delta)=\la c_1, K_1,W_1, d_1 \ra \cdot  \alpha(\delta') \in
         \eta(\alpha(\delta_1) \parallel \alpha(\delta'_2))
         \end{eqnarray}
         and this completes the proof.
\end{description}

\bigskip

\begin{lemma} (Lemma \ref{lem:percomp1})
Let $P$ be a program, $H$ and $G$ be two goals and assume that
$\delta\in  {\cal S}'_P(H,G)$. Then there exists $\delta_1\in
{\cal S}'_P(H)$ and $\delta_2\in  {\cal S}'_P(G)$, such that
$\alpha(\delta) \in \eta (\alpha(\delta_1) \parallel
\alpha(\delta_2))$.
\end{lemma}
{\bf Proof } We construct, by induction on the $l=length(\delta)$
two sequences $\delta \uparrow _{(H,G)}=(\delta_1,\delta_2)$,
where
\begin{enumerate}
    \item for $i=1,2$, $V_{loc}(\delta_i)\subseteq
    V_{loc}(\delta)$
    and $Inc(\delta_i)\subseteq Inc(\delta)$ (and therefore
    $CT \models instore (\delta_i)  \rightarrow instore (\delta)$).
    \item $Ass(\delta_1)\subseteq Ass(\delta) \cup Stable (\delta_2)$
    and $Ass(\delta_2)\subseteq Ass(\delta) \cup Stable(\delta_1)$,
    \item $\delta_1\in {\cal S}'_P(H)$, $\delta_2 \in {\cal S}'_P(G)$,
    $\alpha(\delta_1) \parallel \alpha(\delta_2)$ is defined and
    $\alpha(\delta) \in \eta (\alpha(\delta_1) \parallel \alpha(\delta_2))$.
\end{enumerate}
\begin{description}
    \item[{\bf ($l=1$)}] In this case $\delta= \la (H,G), c, \emptyset,(H,G),c\ra$.
    We define
    \[\delta \uparrow_{(H,G)}=(\la H, c, \emptyset,H,c\ra,
    \la G, c, \emptyset,G,c\ra)=(\delta_1,\delta_2),
    \]
    where $\delta_1 \in {\cal S}'_P(H)$
    and $\delta_2 \in {\cal S}'_P(G)$.
    By definition for $i=1,2$, $V_{loc}(\delta_i)=\emptyset$,
    $Inc(\delta_i)=\{c\}= Inc(\delta)$ and $Ass(\delta_i)=\emptyset$. \\
    Moreover $\alpha(\delta_1)=\la c, \emptyset, H,c\ra $
    and $\alpha(\delta_2)=\la c, \emptyset, G,c\ra$ and then
    $\alpha(\delta_1)\parallel \alpha(\delta_2)$ is defined.
    Now the proof is straightforward by definition of $\parallel$.
    \item[{\bf ($l>1$)}] Assume that $\delta \in {\cal S}'_P(H,G)$. By definition
    $$\delta = \la (H,G), c_1, K_1,B_2,d_1\ra \cdot\delta',$$
    where $\delta'\in {\cal S}'_P(B_2)$ and $t=\la (H,G), c_1, K_1,B_2,d_1\ra $
    is compatible with $\delta'$.
    Recall that, by definition, the tuple $t$
    represents a derivation step
    \[s=\la (H,G), c_1\ra \rrarrow_P^{K_1}
    \la B_2,d_1\ra.\]
    Now we distinguish various cases according
    to the structure of the derivation step $s$.
    \begin{itemize}
        \item In the derivation step $s$, we use the \mbox{\bf Solve'} rule.
         In this case, without loss of generality, we can assume that
         $H=(c,H')$,
         \[s=\la (H,G), c_1\ra \rrarrow_P^{\emptyset}
         \la (H',G),d_1\ra,\]
         $CT \models c_1  \wedge c \leftrightarrow d_1$,
         $t= \la (H,G), c_1, \emptyset,(H',G),d_1\ra$ and
         $\delta' \in {\cal S}'_P(H',G)$. Moreover $\alpha(\delta)=
         \la c_1, \emptyset ,W,d_1\ra \cdot
         \alpha(\delta')$, where $W$ is the first stable multiset of
         $\alpha(\delta')$.\\
         By inductive hypothesis there exist $\delta'_1 \in {\cal S}'_P(H')$ and
         $\delta_2 \in {\cal S}'_P(G)$ such that
         $\delta' \uparrow_{(H',G)}=(\delta'_1,\delta_2)$,
         $\alpha(\delta_1') \parallel
         \alpha(\delta_2)$ is defined and
         $\alpha(\delta') \in \eta(\alpha(\delta_1') \parallel
         \alpha(\delta_2))$.
         Then, we define
         \[\delta \uparrow_{(H,G)}= (\delta_1,\delta_2) \mbox{ where }
         \delta_1= \la H, c_1, \emptyset ,H',d_1\ra \cdot \delta'_1.\]
         By definition $\la H, c_1\ra \rrarrow_P^{\emptyset}
         \la H',d_1\ra$,
         $t'=\la H, c_1, \emptyset ,H',d_1\ra $
         represents a derivation step for $H$,
         $Fv(d_1) \subseteq Fv(H) \cup Fv(c_1)$ and therefore
         $V_{loc}(t')=\emptyset$. Then the following holds.
\begin{enumerate}
   \item Let $i \in [1,2]$. By the inductive hypothesis, by
   construction and by the previous observation
   $V_{loc}(\delta_i) \subseteq V_{loc}(\delta')= V_{loc}(\delta)$ and
   $Inc(\delta_i)\subseteq Inc(\delta') \cup \{c_1\} =Inc(\delta)$.
   \item By inductive hypothesis and by construction, \\
   $\begin{array}{lll}
   Ass(\delta_1)=Ass(\delta'_1)\subseteq Ass(\delta') \cup
   Stable(\delta_2)= Ass(\delta)\cup Stable(\delta_2)
   \mbox{ and }\\
   Ass(\delta_2) \subseteq Ass(\delta') \cup Stable(\delta'_1)
   = Ass(\delta) \cup Stable(\delta_1).
   \end{array}
   $
   \item By inductive hypothesis
   $\delta_2 \in {\cal S}'_P(G)$.
   The proof of the other statements follows by Lemma
   \ref{lem:ingoalH} and  by inductive hypothesis.
\end{enumerate}

     \item In the derivation step $s$, we use the \mbox{\bf Simplify'}
         rule and let us to assume that in the derivation step $s$ atoms
         deriving from $H$ only are rewritten. \\
         In this case, we can assume that
         $s=\la (H,G), c_1\ra \rrarrow_P^{K_1}
         \la (B,G),d_1\ra$, $\delta' \in {\cal S}'_P(B,G)$  and
         $t=\la (H,G), c_1, K_1, (B,G),d_1\ra$.
         By inductive hypothesis there exist $\delta'_1 \in {\cal S}'_P(B)$ and
         $\delta_2 \in {\cal S}'_P(G)$ such that
         $\delta' \uparrow_{(B,G)}=(\delta'_1,\delta_2)$, $\alpha(\delta_1') \parallel
         \alpha(\delta_2)$ is defined and
         $\alpha(\delta') \in \eta(\alpha(\delta_1') \parallel
         \alpha(\delta_2))$.
         Then, we define
         \[\begin{array}{l}
           \delta \uparrow_{(H,G)}= (\delta_1,\delta_2) \mbox{ where }
           \delta_1= \la H, c_1, K_1 ,B,d_1\ra \cdot \delta'_1.
         \end{array}
         \]
         By definition $\la H, c_1\ra \rrarrow_P^{K_1}
         \la B,d_1\ra$, $t'=\la H, c_1, K_1 ,B,d_1\ra $
         represents a derivation step for $H$
         and $ V_{loc}(t')=V_{loc}(t).$

         Now the following holds.
         \begin{enumerate}
         \item Let $i \in [1,2]$. By the inductive hypothesis,
         by construction and by the previous observation
         $V_{loc}(\delta_i)\subseteq V_{loc}(\delta')
         \cup V_{loc}(t)= V_{loc}(\delta)$
         and $Inc(\delta_i)\subseteq Inc(\delta') \cup \{c_1\} =Inc(\delta)$.
         \item By inductive hypothesis and by construction,
         \[
         \begin{array}{llll}
           &Ass(\delta_1)&=& Ass(\delta'_1)\cup  \{K_1\}
           \\
           &&\subseteq&
           Ass(\delta')\cup Stable (\delta_2) \cup
         \{K_1\}=
            Ass(\delta)\cup Stable (\delta_2)\\

           \mbox{and}
           \\
         &Ass(\delta_2)&\subseteq&
           Ass(\delta') \cup Stable(\delta'_1)
         \subseteq
         Ass(\delta)\cup Stable(\delta_1).
        \end{array}\]

         \item By inductive hypothesis $\delta_2 \in {\cal S}'_P(G)$.
   The proof of the other statements follows by Lemma
   \ref{lem:ingoalH} and by inductive hypothesis.
\end{enumerate}

         \item In the derivation step $s$, we use the \mbox{\bf Simplify'}
         rule and let us to assume that in the derivation step $s$ atoms
         deriving both from $H$ and $G$ are rewritten.\\
         In this case, we can assume that $H=(H', H'')$,
         $G=(G',G'')$,
         $H''\neq \emptyset$, $G''\neq \emptyset$,
         $s=\la (H,G), c_1\ra \rrarrow_P^{K_1} \la
         (H',G',B),d_1\ra$,
         $\delta' \in {\cal S}'_P(H',G',B)$ and\\
         $t= \la (H,G), c_1, K_1,
         (H',G',B),d_1\ra$. \\
         By using the same arguments of the previous point there exist
         $\delta'_1 \in {\cal S}'_P(H,G'') $ and
         $\delta'_2\in {\cal S}'_P(G')$ such that
         $\delta \uparrow_{((H,
         G''),G')}=(\delta'_1,\delta'_2)$.\\
         Now, observe that, by Lemma~\ref{lem:dagoalasass}
         and by definition of $\uparrow$, there exists
         $\delta_1 \in {\cal S}'_P(H)$ such that
         $Ass(\delta_1)=Ass(\delta'_1) \cup
         \{G''\}$,
         $\alpha
         (\delta'_1)=\la c_1, K_1, W_1, d_1\ra \cdot \sigma _1$,
         $\alpha
         (\delta_1)=\la c_1, K_1 \uplus \{G''\}, W_1, d_1\ra \cdot \sigma
         _1$ and
         $V(\delta_1)=V(\delta'_1)$ for $V \in \{
         V_{loc}, \, Inc,  \, Stable\}$.\\
         Moreover, since $\delta \in {\cal S}'_P(H,G)$ and $V_{loc}(\delta'_2)
         \subseteq V_{loc}(\delta)$, we have that $Fv(G'') \cap V_{loc}(\delta'_2)=\emptyset$.
         Then by Lemma~\ref{lem:aggoal}, we have that $\delta_2 =\delta'_2
         \oplus \tilde G'' \in {\cal S}'_P(G)$. By construction
         $Stable(\delta_2)=Stable(\delta'_2) \cup
         \{G''\}$ and $V(\delta_2)=V(\delta'_2)$ for $V \in \{
         V_{loc}, \, Inc, \, Ass\}$.

         Then, we define
         \[\begin{array}{l}
         \delta \uparrow_{(H,G)}= (\delta_1,\delta_2).
         \end{array}
         \]
         Now the following holds.
         \begin{enumerate}
          \item
          Let $i \in [1,2]$. By definition of $\uparrow$ and  by the previous observation
         $V_{loc}(\delta_i)= V_{loc}(\delta'_i) \subseteq  V_{loc}(\delta)$
         and $Inc(\delta_i)= Inc(\delta'_i) \subseteq Inc(\delta)$.

         \item By definition of $\uparrow$ and by construction
         $Ass(\delta_1)=Ass(\delta'_1)\cup \{G''\} \subseteq
         Ass(\delta)\cup Stable (\delta'_2)  \cup \{G''\} =
         Ass(\delta)\cup Stable (\delta_2)$ and
         $Ass(\delta_2)=Ass(\delta'_2) \subseteq
         Ass(\delta) \cup Stable (\delta'_1) =
         Ass(\delta)\cup Stable (\delta_1)$.
         \item  The proof that $\alpha(\delta_1) \parallel \alpha
         (\delta_2)$ is defined follows by observing that,
         by definition of derivation, $V_{loc}(\delta'_1 ) \cap
         Fv(G'')=\emptyset$, by
         construction for $i \in [1,2]$,
         $V_{loc}(\delta_i)=V_{loc}(\delta'_i)$ and by definition of
         $\uparrow$, $\alpha(\delta'_1) \parallel \alpha
         (\delta'_2)$ is defined.
        Finally, the proof that $ \alpha(\delta) \in \eta( \alpha(\delta_1) \parallel \alpha
         (\delta_2))$ follows by observing that by definition of
         $\uparrow$,  $ \alpha(\delta) \in \eta( \alpha(\delta'_1) \parallel \alpha
         (\delta'_2))$ and by construction $\eta ( \alpha(\delta'_1) \parallel \alpha
         (\delta'_2)) \subseteq \eta ( \alpha(\delta _1) \parallel \alpha
         (\delta_2))$.
         \end{enumerate}
\end{itemize}
\end{description}

\bigskip

\begin{lemma} (Lemma \ref{lem:percomp2})
Let $P$ be a program, let $H$ and $G$ be two goals and assume that
$\delta_1\in  {\cal S}'_P(H)$ and $\delta_2\in {\cal S}'_P(G)$ are
two sequences such that the following hold:
\begin{enumerate}
    \item $\alpha(\delta_1) \parallel \alpha(\delta_2)$ is defined,
    \item $\sigma= \la c_1, K_1 ,W_1,d_1\ra \cdots
    \la c_n, \emptyset ,W_n,c_n\ra \in
    \eta (\alpha(\delta_1) \parallel \alpha(\delta_2))$,
    \item $(V_{loc}(\alpha(\delta_1)) \cup V_{loc}(\alpha(\delta_2)))
    \cap V_{ass}(\sigma)=\emptyset $,
    \item for $i \in [1,n]$,
    $(V_{loc}(\alpha(\delta_1)) \cup V_{loc}(\alpha(\delta_2)))
    \cap Fv(c_i) \subseteq \bigcup _{j=1} ^ {i-1} Fv(d_j) \cup Fv(W_i)$.
\end{enumerate}
Then there exists $\delta\in {\cal S}'_P(H,G)$ such that
$\sigma=\alpha(\delta)$.
\end{lemma}
{\bf Proof } In the following, given two derivations $\delta_1\in
{\cal S}'_P(H)$ and $\delta_2\in {\cal S}'_P(G)$, which verify the
previous conditions, we construct by induction on the
$l=length(\sigma)$ a derivation $\delta\in {\cal S}'_P(H,G)$ such
that $V_{loc}(\delta)\subseteq V_{loc}(\delta_1) \cup
V_{loc}(\delta_2)$ and $\sigma=\alpha(\delta)$.

\begin{description}
    \item[{\bf ($l=1$)}] In this case $\delta_1=\la H, c,
    \emptyset,H,c\ra$, $\delta_2=\la G, c, \emptyset,G,c\ra$,
    $\alpha(\delta_1)=\la c, \emptyset,H,c\ra$,
    $\alpha(\delta_2)=\la c, \emptyset,G,c\ra$,
    $\sigma=\la c, \emptyset,(H,G),c\ra$
    and $\delta= \la (H,G), c, \emptyset,(H,G),c\ra$.
    \item[{\bf ($l>1$)}]  Without loss of generality, we can assume that
    \[\begin{array}{ll}
      \delta_1=t' \cdot \delta'_1, \
      \delta_2=\la G, e_1, J_1,G_2,f_1\ra \cdot \delta'_2,\\
      \sigma_1=\alpha(\delta_1)=\la c_1, L_1, N_1,d_1\ra \cdot \alpha(\delta'_1)
      \mbox{ and } \\
      \sigma_2=\alpha(\delta_2)=\la e_1, J_1,M_1,f_1\ra \cdot \sigma'_2,\\
    \end{array}
    \]
    where $t'=\la H, c_1, L_1,H_2,d_1\ra$,
    $\delta'_1\in {\cal S}'_P(H_2)$,
    $\sigma \in \eta(\la c_1, L_1, N_1 \uplus M_1,d_1\ra \cdot \bar
    \sigma)$ and
    $\bar \sigma \in \alpha(\delta'_1) \parallel \sigma_2$.

    By definition of $\eta$, there exist the multisets of atoms
    $L',$ $\bar L, $ $L$ and the sequence $\sigma'$ such that
    \[\sigma = \la c_1, L_1 \setminus L,
    ((N_1 \uplus M_1) \setminus \bar L)
    \setminus L',d_1\ra \cdot (\sigma' \setminus L'),
    \]
    where $\sigma'\in \eta(\bar \sigma)\subseteq \eta(\alpha(\delta'_1)\parallel
    \sigma_2)$, $K_1=L_1 \setminus L$ and $W_1=((N_1 \uplus M_1) \setminus \bar L)\setminus
    L'$.
    Now the following holds
    \begin{enumerate}
        \item $\alpha(\delta'_1) \parallel \alpha(\delta_2)$ is
        defined. By definition, we have to prove that
        \[(V_{loc}(\alpha (\delta'_1) )\cup Fv(H_2)) \cap (V_{loc}(\alpha (\delta_2) )\cup
        Fv(G))= Fv(H_2) \cap Fv(G).\]
        First of all, observe that since $V_{loc}(\alpha (\delta'_1)
        )\subseteq V_{loc}(\alpha (\delta_1) )$ and $\alpha(\delta_1) \parallel
        \alpha(\delta_2)$ is defined, we have that $V_{loc}(\alpha (\delta'_1) )
        \cap (V_{loc}(\alpha (\delta_2) )\cup
        Fv(G))= \emptyset$ and $(Fv(H) \cup V_{loc}(\alpha (\delta_1)
        ) \cap (V_{loc}(\alpha (\delta_2))=\emptyset$.

        Now, observe that by definition of
        derivation, $Fv(H_2) \subseteq  Fv(H) \cup V_{loc}(\alpha
        (\delta_1))$.
        Therefore, by previous observations, $Fv(H_2) \cap V_{loc}(\alpha
        (\delta_2))=\emptyset$ and then the thesis.
        \item $\sigma'= \la c_2, K_2 ,W_2 \uplus L',d_2\ra \cdots
        \la c_n, \emptyset ,W_n \uplus L',c_n\ra \in
        \eta (\alpha(\delta'_1) \parallel \alpha(\delta_2))$. The proof is straightforward, by
        definition of $\parallel$.
        \item By definition, by the hypothesis and by
        Lemma~\ref{lem:stessevar}, we have that
        \[\begin{array}{lll}
        (V_{loc}(\alpha(\delta'_1)) \cup V_{loc}(\alpha(\delta_2)))
         \cap V_{ass}(\sigma') & \subseteq \\
         (V_{loc}(\alpha(\delta_1)) \cup
         V_{loc}(\alpha(\delta_2))) \cap
         V_{ass}(\sigma)& = & \emptyset.
         \end{array}
         \]
        \item For $i \in [2,n]$,
        \[\begin{array}{lll}(
        V_{loc}(\alpha(\delta'_1)) \cup V_{loc}(\alpha(\delta_2)))
        \cap Fv(c_i) \subseteq
        \bigcup _{j=2} ^ {i-1} Fv(d_j) \cup Fv(W_i \uplus L').
        \end{array}\]
        To prove this statement observe that by hypothesis and by
        Lemma~\ref{lem:stessevar}, for $i \in [2,n]$,
        \begin{eqnarray}
         && \nonumber ( V_{loc}(\alpha(\delta'_1)) \cup V_{loc}(\alpha(\delta_2)))
        \cap Fv(c_i) \  \subseteq \\
         && \nonumber (V_{loc}(\alpha(\delta_1)) \cup V_{loc}(\alpha(\delta_2)))
        \cap Fv(c_i) \  \subseteq \\
         && \bigcup _{j=1} ^ {i-1} Fv(d_j) \cup Fv(W_i).
          \label{eq:25ott2}
        \end{eqnarray}
        Let $i \in [2,n]$, such that there exists
        $x \in ( V_{loc}(\alpha(\delta'_1)) \cup V_{loc}(\alpha(\delta_2)))
        \cap Fv(c_i) \cap Fv(d_1)$.We have to prove that
        $x \in Fv(W_i)$ and then the thesis.

        First of all, observe that since $x \in Fv(d_1)$, by
        definition of derivation, we have that
        $x \not \in
        V_{loc}(\alpha(\delta'_1))$
        and therefore $x \in  V_{loc}(\alpha(\delta_2))
        \cap Fv(c_i) \cap Fv(d_1)$.

        Moreover, since by hypothesis $\alpha(\delta_1)\parallel \alpha (\delta_2)$ is
        defined, we have that $x \not \in Fv(H) \cup  V_{loc}(t')$.
        Therefore,
        since $x \in Fv(d_1)$ and by definition of derivation, we have
        that $x \in Fv(L_1)\cup Fv(c_1)$. Now we have two
        possibilities
        \begin{itemize}
            \item $x \in Fv(c_1)$. In this case, since
            $x \in V_{loc}(\alpha(\delta_2))$
            and by point 4 of the hypothesis, we have
            that $x \in Fv(W_i)$.
            \item $x \in Fv(L_1)$. In this case there exists
            $A \in L_1$ such that $x \in Fv(A)$. Since by hypothesis
            $(V_{loc}(\alpha(\delta_1)) \cup V_{loc}(\alpha(\delta_2)))
            \cap V_{ass}(\sigma)=\emptyset $, we have that
            $A \not \in Ass(\sigma)$ (i.e. $A \not \in K_1$)
            and therefore, by definition of
            $\parallel$, there exists $A'\in G$ such that
            $CT \models c_1 \wedge A \leftrightarrow c_1\wedge A'$.
            Note that, since $x \in  V_{loc}(\alpha(\delta_2))$, we have
            that $x \not \in Fv(G) \supseteq Fv(A')$. Then $x \in Fv(c_1)$
            and then analogously to the previous case, $x \in Fv(W_i)$.
        \end{itemize}
        Then, by (\ref{eq:25ott2}),
        \[( V_{loc}(\alpha(\delta'_1)) \cup
        V_{loc}(\alpha(\delta_2)))
        \cap Fv(c_i) \ \subseteq \bigcup _{j=2} ^ {i-1} Fv(d_j) \cup Fv(W_i)
        \]
        and then the thesis.
    \end{enumerate}
By previous results and by inductive hypothesis, we have that
there exists $\bar \delta\in {\cal S}'_P(H_2,G)$ such that
$V_{loc}(\bar \delta)\subseteq V_{loc}(\delta'_1) \cup
V_{loc}(\delta_2)$ and  $\sigma'=\alpha(\bar \delta)$. Moreover by
definition of $\eta$, $L'\subseteq (H_2,G)$ is a multiset of atoms
which are stable in $\bar \delta$. Then $ \delta'=\bar
\delta\ominus L' \in {\cal S}'_P(B)$, where the goal $B$ is
obtained from the goal $(H_2,G)$ by deleting the atoms in $L'$. By
construction
\begin{equation}\label{eq:9marzo}
    V_{loc}(\delta')= V_{loc}(\bar \delta) \mbox{ and }
    V_{ass}(\delta')= V_{ass}(\bar \delta).
\end{equation}

Now observe that since $t'=\la H, c_1, L_1,H_2,d_1\ra$ represents
a derivation step for $H$, we have that $t=\la (H,G), c_1,
K_1,B,d_1\ra$ represents a derivation step for $(H,G)$. Let us
denote by $\delta$ the sequence $t \cdot  \delta'$.

Then, to prove the thesis, we have to prove that
$V_{loc}(\delta)\subseteq V_{loc}(\delta_1) \cup
V_{loc}(\delta_2)$, $t$ is compatible with $\delta'$ (and
therefore $\delta \in {\cal S}'_P(H,G)$) and
$\sigma=\alpha(\delta)$.\\

\begin{description}
\item[($V_{loc}(\delta)\subseteq V_{loc}(\delta_1) \cup
V_{loc}(\delta_2)$).]
\[
\begin{array}{lll}
  V_{loc}(\delta) & = & \mbox{by construction} \\
  V_{loc}(t)\cup V_{loc}(\delta') &
  = & \mbox{by (\ref{eq:9marzo})} \\
  V_{loc}(t')\cup V_{loc}(\bar \delta) &
  \subseteq & \mbox{by inductive hypothesis} \\
  V_{loc}(t')\cup V_{loc}(\delta'_1) \cup V_{loc}(\delta_2)&
  = & \mbox{by construction} \\
  V_{loc}(\delta_1) \cup V_{loc}(\delta_2)
\end{array}
\] and then the thesis.
    \item[($t$ is compatible with $\delta'$).] The following
    holds.
    \begin{enumerate}
    \item $CT \models instore(\delta') \rightarrow d_1$.
    The proof is straightforward, since by construction either
    $instore(\delta')=instore(\delta'_1)$ or
    $instore(\delta')=instore(\delta_2)$.
    \item $V_{loc}(\delta') \cap Fv (t) = \emptyset$.
    By construction, (\ref{eq:9marzo}) and by inductive hypothesis
    \begin{eqnarray}
    &&\nonumber
    V_{loc}(t)=V_{loc}(t'), \
    Fv(t) = Fv(t') \cup Fv(G) \mbox{ and } \\
    &&\label{eq:29maggio1}
    V_{loc}(\delta')\subseteq V_{loc}(\delta'_1) \cup
    V_{loc}(\delta_2).
    \end{eqnarray}
    By definition of derivation and since
    $\alpha(\delta'_1) \parallel \alpha(\delta_2)$ is defined,
    we have that
    $V_{loc}(\delta'_1) \cap (Fv (t') \cup Fv(G)) =\emptyset$
    and therefore by the second statement in
    (\ref{eq:29maggio1})
    \begin{equation}\label{eq:29maggio2}
    V_{loc}(\delta'_1) \cap Fv (t) =\emptyset.
    \end{equation}

    By point 3 of the hypothesis
    $Fv(K_1) \cap V_{loc}(\delta_2)=\emptyset$.
    Moreover, since by definition of $\alpha$ and $\parallel$,
    $W_1 \subseteq (H,G)$, we have that
    \[\begin{array}{lll}
      Fv(c_1) \cap V_{loc}(\delta_2) & \subseteq & \mbox{(by point 4 of the hypothesis)} \\
      Fv(W_1) \cap V_{loc}(\delta_2) & \subseteq & \mbox{(by the previous observation)} \\
      Fv(H,G) \cap V_{loc}(\delta_2) & = & \mbox{(by definition of
      derivation and }\\
      && \mbox{ since $\alpha(\delta_1) \parallel \alpha(\delta_2)$ is
    defined)} \\
    \emptyset
    \end{array}
   \]
    Finally, since
    $\alpha(\delta_1) \parallel \alpha(\delta_2)$ is
    defined we have that
    $(Fv(H) \cup V_{loc}(t')) \cap V_{loc}(\delta_2) =\emptyset$.
    Then by definition and by (\ref{eq:29maggio1})
    \begin{eqnarray}
    Fv(t)\cap V_{loc}(\delta_2)& = & (Fv(c_1,H, K_1) \cup V_{loc}(t'))\cap
    V_{loc}(\delta_2)= \emptyset.\label{eq:29maggio3}
\end{eqnarray}
Then
\[\begin{array}{lll}
  V_{loc}(\delta') \cap Fv (t)
& \subseteq  & \mbox{(by the last statement in (\ref{eq:29maggio1}))} \\
  (V_{loc}(\delta'_1) \cup
    V_{loc}(\delta_2)) \cap Fv (t) & \subseteq  & \mbox{(by (\ref{eq:29maggio2}))} \\
   V_{loc}(\delta_2) \cap Fv (t) &  = & \mbox{(by (\ref{eq:29maggio3}))} \\
  \emptyset. \\
\end{array}
\]

    \item $V_{loc}(t)\cap V_{ass}( {\delta'}) = \emptyset$.
    The proof is immediate by the second statement of (\ref{eq:9marzo}),
    since $\sigma'= \alpha (\bar \delta)$, $V_{ass}(\sigma' ) \subseteq
    V_{ass}(\sigma)$, by the first statement in
    (\ref{eq:29maggio1}), since $V_{loc}(t') \subseteq
    V_{loc}(\delta_1)$ and by point 3 of the hypothesis.

    \item for $i \in [2,n]$, $V_{loc}(t) \cap Fv(c_i) \subseteq
    \bigcup_{j=1}^{i-1}  Fv(d_j) \cup V_{stable}( {\delta'})$.
    By construction, since $\delta'=\bar \delta\ominus L' $,
    $\sigma'=\alpha (\bar \delta)$ and $Stable (\sigma')=
    W_n \uplus L'$, we have that $Stable (\delta')=
    W_n$. Then the proof is immediate by observing that
    $V_{loc}(t)=V_{loc}(t') \subseteq V_{loc}(\delta_1)$, for
    $i \in [2,n]$, $W_i \subseteq W_n$ and by point 4 of the hypothesis.
\end{enumerate}
    \item[($\sigma=\alpha(\delta)$).] By inductive hypothesis
    $\sigma'=\alpha(\bar \delta)$ and then by construction
    $\sigma'\setminus L'=\alpha(\delta').$ Then
    \[
      \sigma = \la c_1, K_1,
    W_1,d_1\ra \cdot (\sigma' \setminus L') =  \la c_1, K_1,
    W_1,d_1\ra \cdot \alpha(\delta')=  \alpha(\delta),
    \]
    where the last equality follows by observing that
    $\delta=t\cdot \delta' $, where
    \[ t=\la (H,G), c_1, K_1,B,d_1\ra\]
    and $W_1$ is the multiset of all the atoms in $(H,G)$, which are stable in $\delta$.
\end{description}
\end{description}

\end{document}